\documentclass[aps,twocolumn,nofootinbib,floatfix,superscriptaddress]{revtex4}
\usepackage{amsfonts}
\usepackage{amssymb}
\usepackage{amsmath}
\usepackage{graphics}
\usepackage{graphicx}
\usepackage{soul}
\usepackage{natbib}
\usepackage{bm}
\usepackage{mathtools}
\usepackage{stmaryrd}
\usepackage{epstopdf}
\usepackage{color}
\usepackage{lipsum}
\usepackage{balance}
\usepackage{enumitem}
\usepackage[colorlinks=true,citecolor=black,linkcolor=black,urlcolor=black,filecolor=black]{hyperref}
\usepackage[dvipsnames]{xcolor}
\usepackage{bbold}
\usepackage{url}

\usepackage{acronym}
\newacro{VRR}{Vector Resonant Relaxation}
\newcommand{\VRR}{\ac{VRR}}
\newacro{DIA}{Direct Interaction Approximation}
\newcommand{\DIA}{\ac{DIA}}
\newacro{DF}{distribution function}
\newcommand{\DF}{\ac{DF}}
\newacro{MSR}{Martin--Siggia--Rose}
\newcommand{\MSR}{\ac{MSR}}
\newacro{FDT}{Fluctuation-Dissipation Theorem}
\newcommand{\FDT}{\ac{FDT}}
\newacro{BH}{black hole}
\newacroplural{BH}[BHs]{black holes}
\newcommand{\BH}{\ac{BH}}
\newcommand{\BHs}{\acp{BH}}

\newcommand{\p}{\partial}
\newcommand{\rd}{\mathrm{d}}
\newcommand{\re}{\mathrm{e}}
\newcommand{\ri}{\mathrm{i}}
\newcommand{\bK}{\mathbf{K}}
\newcommand{\bL}{\mathbf{L}}
\newcommand{\hbL}{\widehat{\mathbf{L}}}

\newcommand{\Htot}{H_{\mathrm{tot}}}
\newcommand{\mH}{\mathcal{H}}
\newcommand{\bKp}{\mathbf{K}^{\prime}}
\newcommand{\vphid}{\varphi_{\mathrm{d}}}
\newcommand{\vphi}{\varphi}
\newcommand{\deltaD}{\delta_{\mathrm{D}}}
\newcommand{\half}{\tfrac{1}{2}}
\newcommand{\Tc}{T_{\mathrm{c}}}
\newcommand{\G}{\Gamma}
\newcommand{\mJ}{\mathcal{J}}
\newcommand{\ml}{\langle}
\newcommand{\mr}{\rangle}
\newcommand{\hvphi}{\widehat{\varphi}}

\newcommand{\eps}{\epsilon}
\newcommand{\1}{\mathbf{1}}
\newcommand{\2}{\mathbf{2}}
\newcommand{\3}{\mathbf{3}}
\newcommand{\4}{\mathbf{4}}
\newcommand{\5}{\mathbf{5}}
\newcommand{\6}{\mathbf{6}}
\newcommand{\bn}{\mathbf{n}}
\newcommand{\vsigma}{\varsigma}
\newcommand{\tmin}{\scalebox{0.6}{$-$}}
\newcommand{\tplus}{\scalebox{0.6}{$+$}}
\newcommand{\bM}{\mathbf{M}}
\newcommand{\ellint}{\ell_{\mathrm{int}}}
\newcommand{\Gz}{g}
\newcommand{\bX}{\mathbf{X}}
\newcommand{\Min}{\mathrm{Min}}
\newcommand{\Max}{\mathrm{Max}}
\newcommand{\Ht}{H_{\mathrm{test}}}
\newcommand{\MBH}{M_\bullet}
\newcommand{\bO}{\mathbf{\Omega}}
\newcommand{\mO}{\mathcal{O}}
\newcommand{\hbO}{\widehat{\mathbf{\Omega}}}
\newcommand{\hR}{\widehat{R}}
\newcommand{\tR}{\widetilde{R}}
\newcommand{\tr}{\widetilde{r}}
\newcommand{\ellmax}{\ell_{\mathrm{max}}}
\newcommand{\nmax}{n_{\mathrm{max}}}

\newcommand{\rG}{\mathrm{G}}
\newcommand{\test}{\mathrm{test}}
\newcommand{\rms}{\mathrm{rms}}
\newcommand{\mR}{R}
\newcommand{\Etot}{E_{\mathrm{tot}}}

\usepackage{soul} \usepackage{xcolor}

\DeclareRobustCommand{\bf}[1]{{\textbf{#1}}}
\DeclareRobustCommand{\u}[1]{{\underline{#1}}}

\begin{document}

\title{Vector Resonant Relaxation and Statistical Closure Theory.
\\
I. Direct Interaction Approximation}

\author{Sofia Flores}
\affiliation{Institut d'Astrophysique de Paris, UMR 7095, 98 bis Boulevard Arago, F-75014 Paris, France}
\author{Jean-Baptiste Fouvry}
\affiliation{Institut d'Astrophysique de Paris, UMR 7095, 98 bis Boulevard Arago, F-75014 Paris, France}

\begin{abstract}
Stars orbiting a supermassive black hole
in the center of galaxies undergo very efficient
diffusion in their orbital orientations: this is ``Vector Resonant Relaxation''.
Such a dynamics is intrinsically non-linear, stochastic, and correlated,
hence bearing deep similarities with turbulence in fluid mechanics or plasma physics.
In that context, we show how generic methods stemming
from statistical closure theory,
namely the celebrated ``Martin--Siggia--Rose formalism'',
can be used to characterize the correlations
describing the redistribution of orbital orientations.
In particular, limiting ourselves to the leading order truncation in this closure scheme,
the so-called ``Direct Interaction Approximation'',
and placing ourselves in the limit of an isotropic distribution of orientations,
we explicitly compare the associated prediction
for the two-point correlation function
with measures from numerical simulations.
We discuss the successes and limitations of this approach
and present possible future venues.
\end{abstract}
\maketitle

\section{Introduction}
\label{sec:Introduction}

Most galaxies contain a supermassive \BH\ in their center~\citep{Kormendy+2013}. Recent observations keep offering us new information
on these galactic behemoths, in particular on SgrA*,
the supermassive \BH\ in the center of the Milky-Way.
This includes in particular
(i) a detailed census of the stellar populations therein~\citep{Ghez+2008,Gillessen+2017}
highlighting the presence of a clockwise stellar disc~\citep[see, e.g.\@,][]{Paumard+2006};
(ii) the observation of a cold accretion disc~\citep{Murchikova+2019};
(iii) the observation of the relativistic precession of the star S2~\citep{Gravity+2020};
(iv) the observation of SgrA*'s horizon shadow~\citep{EHT2022}.
All these recent successes call
for the development of appropriate theoretical frameworks
to interpret observations, in particular regarding the statistical
distribution of the stellar orbits.

In practice, the long-term evolution of stars around
a supermassive \BH\ involves a wealth of dynamical processes~\citep{Rauch+1996,Merritt2013,Alexander2017}.
Here, we focus on the process of \VRR\@~\citep{Rauch+1996,Gurkan+2007,Eilon+2009,Kocsis+2015},
the mechanism that drives the efficient diffusion of the stellar orbital orientations
through coherent resonant torques between the orbits.
Astrophysically, this process is particularly important
to understand the warping of the stellar disc around SgrA*~\citep[see, e.g.\@,][]{Kocsis+2011},
to constrain the efficiency of binary mergers
in this dense stellar environment~\citep{Hamers+2018},
or to investigate possible discs of intermediate mass~\BHs\@~\citep{Szolgyen+2018},
to name a few.

\VRR\ is an archetype of long-range dynamics,
just like it occurs in plasmas~\citep{Nicholson1992},
self-gravitating clusters~\citep{Binney+2008},
or in more generic class of systems~\citep{Campa+2014}.
More precisely, in \VRR\@, the instantaneous orbital orientation
of each star is tracked via a single unit vector, its normalized angular momentum vector.
The angular momentum vector can then be interpreted as a massive particle evolving on the unit sphere, with the vector's norm analoguous to its ``mass''.
Accordingly, in the right set of canonical coordinates,
phase space is the unit sphere~\cite{Kocsis+2015}.
This is formally similar to the dynamics of classical Heisenberg spins~\citep[see, e.g.\@,][]{Gupta+2011},
or the Maier--Saupe model for liquid crystals~\citep{Maier+1958,Roupas+2017}.
Importantly, within the \VRR\@ model, individual particles on the sphere 
have no ``kinetic energy''
following the orbit-average of the system's Hamiltonian~\citep{Kocsis+2015,Roupas2020}.
This is a feature shared with other important models of long-range interacting systems,
such as plasma diffusion in two dimensions in the presence of a strong magnetic field~\citep[see, e.g.\@,][]{Taylor+1971}
or the relaxation of point vortices in two-dimensions~\citep[see, e.g.\@,][]{Chavanis+1996}.
In \VRR\@, particles are then coupled to one another
via a long-range pairwise interaction potential,
whose precise spectrum depends on the considered orbits~\citep[see appendix~{B} in][]{Kocsis+2015}.
Because the gravitational couplings are pairwise,
the system's evolution equation is then naturally quadratically non-linear.
Given all these elements,
it is of no surprise that \VRR\ drives inevitably
a long-range, non-linear and stochastic dynamics.

To get a better grasp on \VRR\@,
one is therefore interested in characterizing
the (ensemble-)averaged correlation functions of the system's fluctuations.
These correlations are said to be spatially extended~\citep{GarciaOjalvo+1999}
in the sense that they non-trivially depend on both
the positions on the sphere and the considered times.
Of prime importance is the two-point correlation function
on which we focus in this work.
Estimating this correlation function from first principles
is no easy task. Indeed, one is unavoidably faced
with the problem of statistical closure,
a difficulty that traverses the characterization
of turbulence in plasma physics~\citep[see][for a review]{Krommes2002}
and fluid dynamics~\citep[see][for a review]{Zhou2021}.

A first milestone in formally approaching this problem was made in~\cite{Kraichnan1959}
that introduced the so-called \DIA\@.
In practice, this approach leads to a set of two non-linear partial integro-differential equations coupling the system's two-point correlation function
and its average response function (that describes the system's response
to infinitesimal fluctuations).
A second seminal milestone
was presented in~\cite{MSR1973},
which developed a generating-functional formalism and renormalization techniques
to derive self-consistent approximations of correlation functions.
We refer to~\cite{Krommes2002} for a detailed historical account
of all these works.
Importantly, this \MSR\ formalism provides an elegant unification of previous approaches.
Indeed, the \DIA\ naturally appears as the leading order approximation of the \MSR\@ scheme.
This is the venue on which we focus here.
We show how the \VRR\ dynamics
naturally lends itself to the \MSR\ formalism.
Furthermore, we show how the \DIA\ can be explicitly implemented
for that system, and compare it with detailed numerical simulations.

The paper is organized as follows.
In Section~\ref{sec:VRR}, we present the main equations of \VRR\@.
In Section~\ref{sec:MSR}, we (briefly) review
the key steps of the \MSR\ formalism,
and how it leads to the \DIA\ closure at leading order.
In Section~\ref{sec:Application},
we explicitly apply this approach to \VRR\
and compare with numerical simulations.
Finally, we conclude in Section~\ref{sec:Conclusion}.
Throughout the main text, technical details are kept to a minimum
and deferred to Appendices or to relevant references.

\section{Vector Resonant Relaxation}
\label{sec:VRR}

We are interested in the process of \VRR\@~\citep{Kocsis+2015}. 
We consider a system of ${ N \!\gg\! 1 }$ stars
orbiting a supermassive black hole (BH) of mass $M_\bullet$, where (i) the total stellar mass is significantly less than the mass of the black hole; (ii) each star follows a quasi-Keplerian precessing orbit; (iii) timescales are longer than the orbital time and the in-plane precession, but shorter than the diffusion time in orbital eccentricity and semi-major axis; (iv) the reorientation of orbital planes is mainly driven by coherent torques between the stellar orbits.

\begin{figure}[htbp!]
	\begin{center}
	\includegraphics[height=0.15\textwidth]{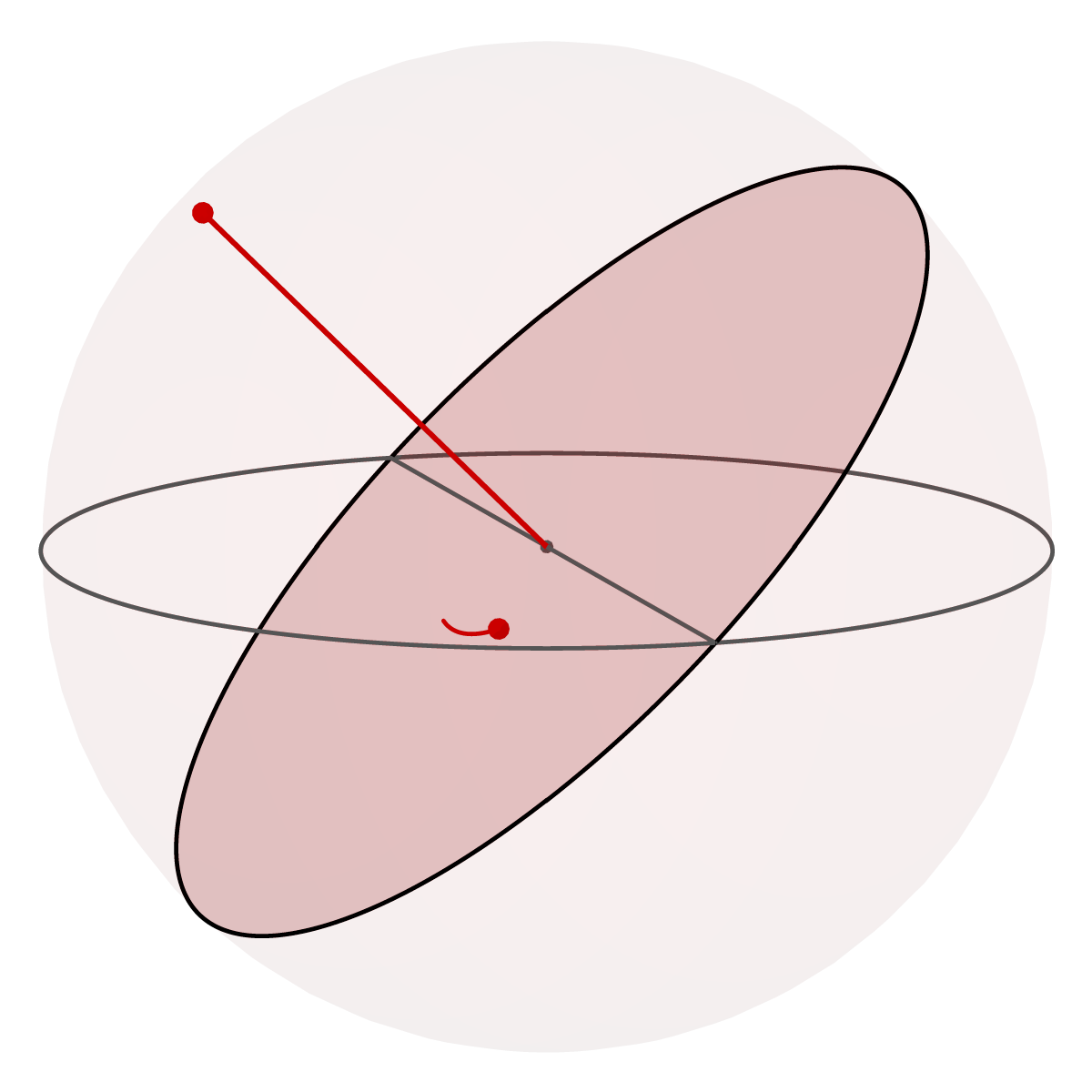}
	\hspace{0.3em}
	\includegraphics[height=0.15\textwidth]{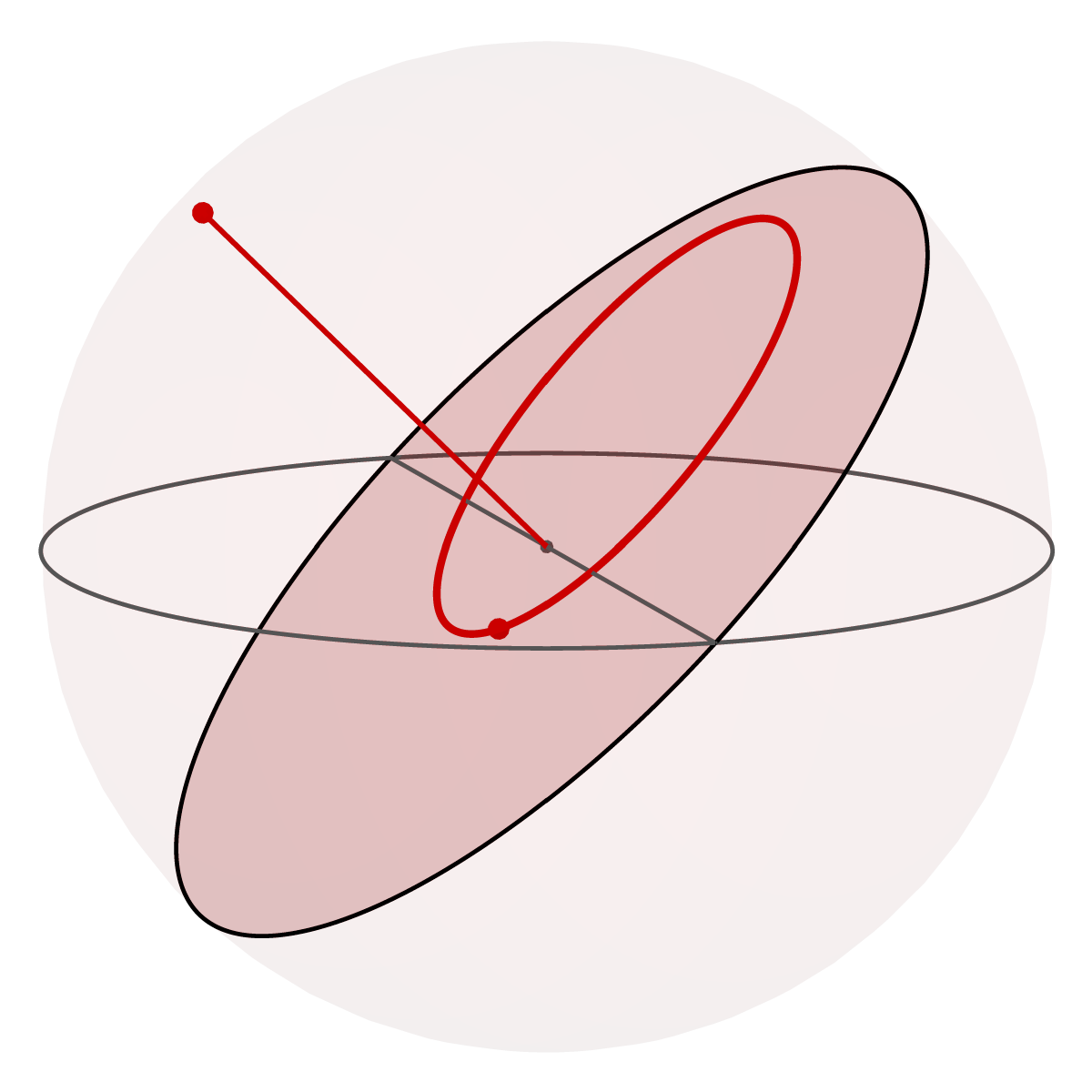}
	\hspace{0.3em}
	\includegraphics[height=0.15\textwidth]{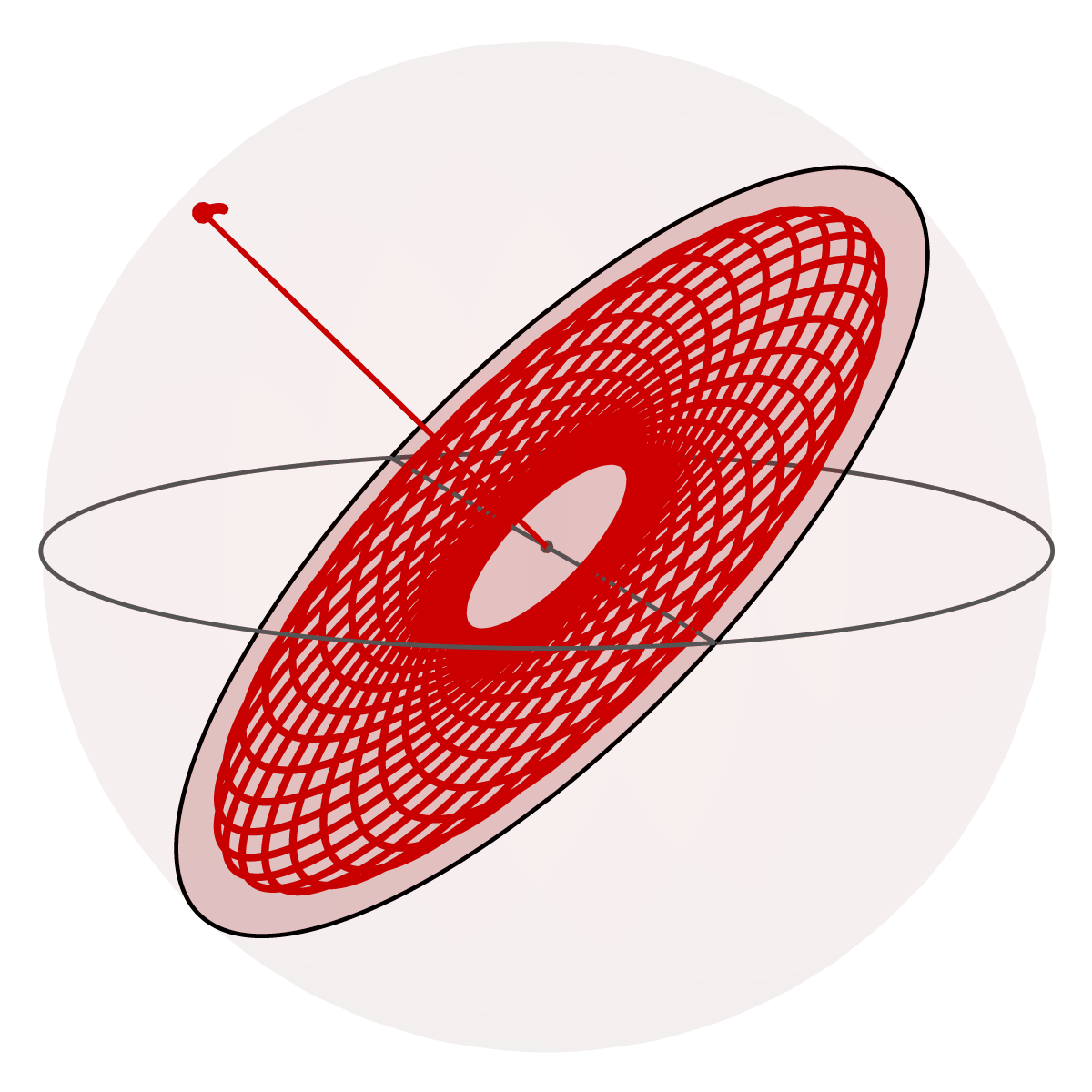}
	\caption{\small{Illustration of the averaging process leading to the \VRR\ equations of motion.
		For the star S2 around SgrA*, we typically have~\citep{Kocsis+2011} (from left to right): orbital motion $\sim$10 years, pericenter precession $\sim$30,000 years, orbital plane reorientation $\sim$1,000,000 years.}}
	\label{fig:wire_to_annuli}
	\end{center}
\end{figure}

Due to the different timescales involved, one can perform a double orbit-average over the short-term evolutions, that is an average over both the Keplerian orbits and the in-plane precession angles.
Following the first average, the resulting Hamiltonian
represents the interaction between two elliptical wires.
With the second average, the eccentric wires become axisymmetric annuli \cite{Kocsis+2015}, see Fig.~\ref{fig:wire_to_annuli}. 
The initial average over the Keplerian motion implies the conservation of the semimajor axis $a$,
while the second average results in the conservation of the eccentricity $e$.
The shape of each annulus is then described
by the conserved quantities
${ \bK \!=\! (m, a, e) }$, with $m$ the stellar mass.
As a result, the norm of each star's angular momentum,
${ L (\bK) \!=\! m \sqrt{G M_\bullet a (1 \!-\! e^{2})} }$, is also conserved.
We point out that (i) the Schwarzschild relativistic in-plane precession from the \BH\@
is naturally accounted for through averaging;
(ii) the Lense-Thirring relativistic out-of-plane precession
from a spinning \BH\@ is neglected.

In \VRR\@, the only dynamical quantity is the orbital orientation,
which we track via the unit vector, $\hbL$, with ${ \bL(\bK,t) \!=\! L(\bK) \hbL(t) }$. 
The study of \VRR\ reduces then to examining the long-term evolution of each star's vector $\hbL$.
The dynamics of the system is therefore simplified to the dynamics of $N$ particles on the unit sphere,
where each particle is labelled by $L(\bK)$ and represents
the instantaneous orientation of an orbital plane associated with a star, see Fig.~\ref{fig:sphere}.
In this set-up, phase space is the unit sphere (see Appendix~\ref{app:Equations_motion}).
\begin{figure}[htbp!]
	\begin{center}
		\includegraphics[width=0.22\textwidth]{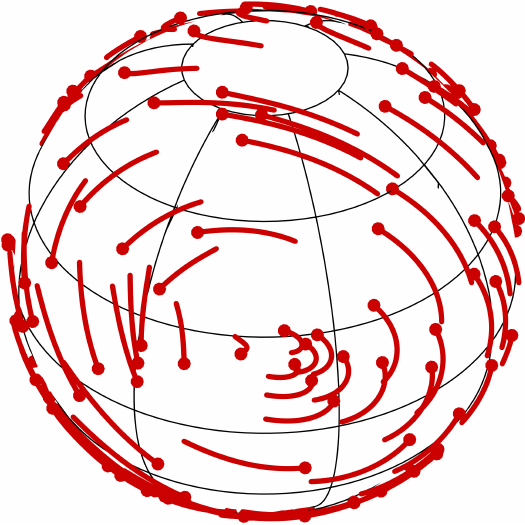}
		\caption{\small{Illustration of \VRR\@,
		namely $N$ particles evolving on the unit sphere,
		where each particle represents the orientation of a star's orbital plane. The tail illustrates the past trajectory of a particle over a finite time interval.}}
		\label{fig:sphere}
	\end{center}
\end{figure}

As derived in~\citep{Kocsis+2015, Magnan+2022}, after a Legendre expansion of the Newtonian interaction and the use of the addition theorem for spherical harmonics,
the \VRR\ total Hamiltonian reads
  \begin{equation}
	\Htot = - \sum_{i < j}^{N} \sum_{\ell = 0}^{+ \infty} \sum_{m = - \ell}^{\ell} \mH_{\ell} [\bK_{i} , \bK_{j}] \, Y_{\ell m} (\hbL_{i}) \, Y_{\ell m} (\hbL_{j}) .
	\label{rewrite_Htot}
\end{equation}
In that expression, the real spherical harmonics ${ Y_{\ell m} (\hbL) }$ are normalized
so that ${ \!\int\! \rd \hbL Y_{\ell m} Y_{\ell' m'} \!=\! \delta^{\ell'}_{\ell} \delta^{m'}_{m} }$,
with ${ \delta^{\ell'}_{\ell} }$ the usual Kronecker symbol.
The isotropic coupling coefficients ${ \mH_{\ell} [\bK , \bKp] }$ are spelled out in Appendix~\ref{App:CouplingCoefficients}.
The equations of motion generated by Eq.~\eqref{rewrite_Htot}
are also detailed in Appendix~\ref{app:Equations_motion}.

Following~\cite{Fouvry+2019},
for a given realization,
we introduce the empirical \DF\@ at time $t$ via
\begin{equation}
\vphid (\hbL , \bK , t) = \sum_{i = 1}^{N} \deltaD \big[ \hbL \!-\! \hbL_{i} (t) \big] \, \deltaD \big[ \bK \!-\! \bK_{i} \big ] ,
\label{def_Fd}
\end{equation}
with $\deltaD$ the usual Dirac delta.
The empirical \DF\@ satisfies the normalization convention
${ \!\int\! \rd \hbL \rd \bK \, \vphid \!=\! N }$.
Here, $\vphid$ satisfies the continuity equation
\begin{equation}
	\p_t \vphid (\hbL , \bK, t)= - \p_{\hbL} \!\cdot\! \big[ \vphid(\hbL , \bK, t) \, \rd \hbL / \rd t \big] ,
\label{evol_vphid}
\end{equation}
with ${ \rd \hbL / \rd t }$ given by Eq.~\eqref{EOM_ind}. To capture the system's rotational invariance, we expand the empirical \DF\@ in spherical harmonics as follows
\begin{equation}
	\vphid (\hbL , \bK , t) = \sum_{a} \vphi_{a} (\bK , t) \, Y_{a} (\hbL) ,
	\label{expansion_vphid}
\end{equation}
with ${ a \!=\! (\ell,m) }$, not to be confused
with the mass and semi-major axis previously introduced. 
The harmonic components ${ \vphi_{a} (\bK , t) }$ are used as tracers of the correlated dynamics occurring in the system.
In the context of \VRR, the number $\ell$ is formally analogous to the wave vector ${k\!=\!|\bf{k}|}$ in plasma turbulence~\citep[see, e.g.\@,][]{Krommes2002},
i.e.\ it describes the typical (inverse) scale of the considered fluctuations.

By expanding in spherical harmonics, Eq.~\eqref{evol_vphid} ultimately becomes (see eq.~{9} in~\cite{Fouvry+2019})
\begin{align}
	\p_t \vphi_{a} (\bK, t) {} & = \half \sum_{b , c} \!\! \int \!\! \rd \bK_{b} \rd \bK_{c} \, \gamma_{a b c} [\bK, \bK_{b},\bK_{c}] \, 
	\nonumber
	\\
	{} & \times \vphi_{b} (\bK_{b} , t) \, 	\vphi_{c} (\bK_{c} , t) ,
	\label{eq:Master_Equation}
\end{align}
where we introduced the time-independent non-random \textit{bare interaction coefficient}, $\gamma_{a b c}$,
explicitly given in Appendix~\ref{app:Interaction_coeff}.
This coefficient captures the coupling between different populations and scales, as well as the system's spherical symmetry.
Equation~\eqref{eq:Master_Equation} is the master equation
on which all the upcoming calculations will be focused.

We now introduce a generalized coordinate that includes time,
${ 1 \!=\! (a, \bK , t) }$.
Equation~\eqref{eq:Master_Equation} becomes
\begin{equation}
\p_{t} \vphi_1 = \half \, \gamma_{1 2 3} \, \vphi_2 \, \vphi_3 ,
\label{eq:Master_Equation_short}
\end{equation}
where ${ \vphi_{a} (\bK, t) \!=\! \vphi_1 }$,
and integration/summation over repeated variables is assumed. 
Here, the generalized bare interaction coefficient is symmetric in its last two arguments, i.e.\ ${\gamma_{1 2 3} \!=\! \gamma_{1 3 2}}$, and its time dependence is a Dirac delta function (see Appendix~\ref{app:Interaction_coeff}).

Equation~\eqref{eq:Master_Equation_short}, the evolution equation for $\vphi_{1}$, is deterministic:
stochasticity only enters through the initial conditions.
In addition, Eq.~\eqref{eq:Master_Equation_short} is local in time,
and its quadratic behavior is very general.
With the addition of constant and linear terms,
this equation can describe turbulence in fluids~\citep[see, e.g.\@, eq.~(1) in][]{Berera+2013} or plasmas~\citep[see, e.g.\@, eq.~(2.5) in][]{Krommes2015}, as well as the growth of large scale cosmological structures~\citep[see, e.g.\@, eq.~(15) in][]{Bernardeau+2012}.

To characterize the correlated stochastic dynamics of the system, our goal is to investigate the statistical properties of the density fluctuations, $\vphi_{1}$, generated by the $N$ particles, recalling that ${ N \!\gg\! 1 }$.
More precisely, we focus on studying the correlation functions of these density fluctuations. In that view, we define the two-point correlation function
\begin{align}
\label{eq:def_C}
C_{1 2} = \ml \vphi_1 \, \vphi_2 \mr - \ml \vphi_1 \mr \ml \vphi_2 \mr,
\end{align}
where the ensemble average, ${ \ml \cdot \mr }$,
is taken over the set of initial conditions.
For a (statistically) isotropic system,
the mean field is ${ \ml \vphi_1 \mr \!=\! 0 }$,
as long as ${ (\ell_1,m_1) \!\neq\! (0,0) }$.
Hence, focusing our interest on isotropic \VRR\@,
we simply have ${ C_{1 2} \!=\! \ml \vphi_1 \, \vphi_2 \mr }$.
Let us note that ${ C \!=\! C_{1 2} }$ is an Eulerian correlation,
since its space and time coordinates are specified independently.
This differs from its Lagrangian definition,
for which the spatial coordinate would be evaluated
along the time-dependent trajectory.
The goal of this work is to predict $C$.

From Eq.~\eqref{eq:Master_Equation_short},
we can derive an evolution equation for~$C$.
It reads
\begin{align}
	\p_{t}C_{1 2}=  \half \, \gamma_{1 3 4} \, \ml \vphi_2 \, \vphi_3 \, \vphi_4 \mr.
	\label{eq:evol_C}
\end{align}
Equation~\eqref{eq:evol_C} manifestly raises the question of Gaussianity.
For isotropic \VRR\@, fluctuations start mainly Gaussian as a result of the particles' initial statistical independence.
Yet, because one wants ${ \p_{t} C }$ to be non-zero, one must unavoidably take into account the non-Gaussianities arising from the evolution of higher-order cumulants.
From Eq.~\eqref{eq:evol_C}, it stems that the evolution equations for the correlations follow a hierarchy that is not self-contained.
This issue is famously known as the closure problem~\citep[see, e.g.\@,][]{Krommes2002}.
To address this problem, we employ a statistical closure scheme
leading to a self-consistent equation for $C$.

\section{Martin--Siggia--Rose closure}
\label{sec:MSR}

To obtain a time-evolution equation for the two-point correlation function, we follow the \MSR\ formalism~\cite{MSR1973}.
Using techniques from Quantum Field Theory~\citep[see, e.g.\@][]{Lancaster+2014,Peskin+2018},
this approach aims at obtaining a renormalized statistical theory
for a classical field satisfying a non-linear dynamical equation, such as Eq.~\eqref{eq:Master_Equation_short}.
We refer to~\cite{Krommes2002} for a particularly extensive and thorough review of the topic of statistical closure theory. Specifically, the present section closely follows section~{6.2} of~\cite{Krommes2002},
whose main details are reproduced below for completeness.

\subsection{Field equations of motion}

To build a statistical description of the system,
it is intuitive to consider both the system's correlations
and its response to infinitesimal fluctuations.
This will take the form of self-consistent equations
involving the two-point correlation and response.

We introduce the mean response function ${R\!=\!R_{1 2}}$,
describing the mean response of the system $\vphi$ at point $1$ to a perturbation $\eta$ at point $2$.
As a result of causality, it reads
\begin{equation}
	\label{eq:def_R_response_main}
	R_{1 2} = \ml \delta \vphi_1 / \delta \eta_{2} \mr \big|_{\eta = 0} \, , \quad \forall \, t_1 > t_2.
\end{equation}
Here, $\eta$ is as a non-random external source added to the equation of motion~\eqref{eq:Master_Equation_short}.
Although explicit, the previous definition does not allow for the derivation
of the desired evolution equations, as we want to combine the dynamics of $C$ and $R$ into a single equation.
Therefore, an alternative representation of the response is required,
i.e.\ one that gives $R$ the same structure as $C$.
This is what the \MSR\ formalism provides.
To do so, \MSR\ proceeds by introducing a new operator, $\hvphi$,
that does not commute with the original field, $\vphi$. A heuristic definition would be ${ \hvphi \!=\! - \delta / \delta \vphi }$. Importantly,
as detailed in Appendix~\ref{app:Response},
with the appropriate choice of $\hvphi$,
Eq.~\eqref{eq:def_R_response_main} is equivalent to
\begin{equation}
\label{eq:Response_time}
R_{1 2} = \ml \vphi_1 \, \hvphi_2 \mr -  \ml \vphi_1 \mr \ml \hvphi_2 \mr,  \quad \forall \, t_1 > t_2 ,
\end{equation}
to compare with Eq.~\eqref{eq:def_C}.

From Eq.~\eqref{eq:Master_Equation_short},
one can then show that $\hvphi$ evolves through
${ \p_{t_1} \hvphi_1 \!=\! - \gamma_{2 3 1} \, \hvphi_2 \, \vphi_3 }$.
To combine the evolution equations for $\vphi$ and $\hvphi$,
\MSR\ introduces the spinor indices
${ \eps \!=\! \pm }$, such that ${ \1 \!=\! (\eps, 1) }$.
From it, one then defines the two-component field,
${ \Phi(+,1) \!=\! \Phi^{\tplus}_1 \!=\! \vphi_1 }$ and ${ \Phi^{\tmin}_1 \!=\! \hvphi_1 }$.
The equation of motion for $\Phi$ ultimately takes
the compact form
(see eq.~{3.8} in~\cite{MSR1973} and eq.~{275} in~\cite{Krommes2002})
\begin{align}
\p_{t_{1}} \Phi_\1 = \half \, \vsigma_{\1 \2} \, \gamma_{\2 \3 \4} \, \Phi_\3 \, \Phi_\4 .
\label{eq:evol_Phi}
\end{align}
In that expression, we introduced
${ \vsigma_{\1 \2} \!=\! \vsigma_{\eps_{1} \eps_{2}} \delta_{1 2} }$, with $\vsigma$ a Pauli-like tensor defined in Eq.~\eqref{eq:def_vsigma}, and ${ \delta_{1 2} \!=\! \delta^{a_2}_{a_1} \deltaD (\bK_1\!\!-\! \bK_2)\deltaD(t_1\!\!-\!t_2) }$. 
Equation~\eqref{eq:evol_Phi} involves the generalized non-random
\textit{bare interaction vertex} $\gamma_{\1 \2 \3}$, defined in Appendix~\ref{app:Interaction_vertex}.
This vertex is fully symmetric,
making Eq.~\eqref{eq:evol_Phi} also fully symmetric.

\subsection{Generating cumulants}

To obtain the correlation and response functions from the field $\Phi$,
\MSR\ uses moment and cumulant generating functionals.
Importantly, this derivation requires \textit{Gaussian} initial conditions and \textit{non-random} coupling coefficients~$\gamma$. Such conditions are verified in isotropic \VRR, although a generalization is possible~\citep[see, e.g.\@,][]{Rose1985}. 
In this section, we employ the same notations as in~\cite{Krommes2002}.

The time-ordered cumulant generating functional is
\begin{align}
W[\eta]= \ln \, \ml \exp{(\Phi_\1\eta_\1)} \mr,
\label{eq:def_Z}
\end{align}
where ${ \eta \!=\! (\eta^{\tplus}, \eta^{\tmin}) }$ is a two-component non-random field. In this context, it is $\vsigma_{\1 \2} \eta_{\2}$ that perturbs the evolution equation of $\Phi$, Eq.~\eqref{eq:evol_Phi} .
From Eq.~\eqref{eq:def_Z}, we define the mean field, ${ \ml \vphi \mr }$, the two-point correlation, $C$, and the mean response function, $R$, as
\begin{equation}
	\label{eq:C_R_W}
	\ml \vphi_1 \mr \!=\! \frac{\delta W}{\delta \eta^{\tplus}_1 }\!\bigg|_{\eta=0}\!\!\!,
	\,C_{1 2} \!=\! \frac{\delta^2 W}{\delta \eta^{\tplus}_1 \delta \eta^{\tplus}_2}\!\bigg|_{\eta=0}\!\!\!, 
	\,R_{1 2} \!=\! \frac{\delta^2 W}{\delta \eta^{\tplus}_1 \delta\eta^{\tmin}_2}\!\bigg|_{\eta =0}\!\!\!.
\end{equation}
As detailed in Appendix~\ref{app:Response},
starting from this expression for $R$,
the time-ordering in Eq.~\eqref{eq:def_Z} gives Eq.~\eqref{eq:Response_time}. 

It is further convenient to reason in terms of cumulants, and compute their equations of motion.
For a given $\eta$, cumulants are defined through 
  \begin{equation}
  	\label{eq:def_cumulants}
	G_{\1...\bf{n}} = \frac{\delta^n W[\eta]}{\delta \eta_{\1}...\delta\eta_\bf{n}}.
 \end{equation}
When taking ${ \eta \!=\! 0 }$ in Eq.~\eqref{eq:def_cumulants}, $G_\1$ reduces to the mean field, and $G_{\1 \2}$ to the desired two-point correlation and response functions. More precisely, we have 
\begin{align}
	{} & G^{\tplus \tplus}_{1 2}|_{\eta=0} = C_{1 2}; \quad G^{\tplus \tmin}_{1 2}|_{\eta=0} = R_{1 2}; 
	\nonumber
	\\
	{} & G^{\tmin \tplus}_{1 2}|_{\eta=0} = R_{2 1}; \quad G^{\tmin \tmin}_{1 2}|_{\eta=0} = 0 ,
	\label{eq:def_G}
\end{align}
with the notation ${ G_{\1 \2} \!=\! G^{\scalebox{0.55}{$ \eps_1  \eps_2 $}}_{1 2} }$.
Importantly, $G$ is fully symmetric,
specifically ${ G_{\1 \2} \!=\! G_{\2 \1} }$.
Our goal is now to obtain an evolution equation for $G_{\1 \2}$,
thus for $C$ and $R$.

\subsection{Cumulant evolution equations}
Appendix~\ref{app:mean_field} provides the detailed
derivation of the evolution equations for the one-point and two-point cumulants. 
We first obtain the equation for $G_\1$ by differentiating Eq.~\eqref{eq:def_cumulants} wrt the time $t_1$. Then, from ${ G_{\1 \1'} \!=\! \delta G_\1 / \delta \eta_{\1'} }$ (see Eq.~\ref{eq:def_cumulants}), we derive the equation for $G_{\1 \1'}$ by differentiating ${\p_{t_1}\!G_\1}$ wrt $\eta_{\1'}$. It reads
\begin{equation}
	\label{eq:dt_G_2_short}
	G_{\1 \1'} =  \Gz_{\1 \1'} + \half \Gz_{\1 \2} \gamma_{\2 \3 \4} G_{\3 \4 \1'},
\end{equation}
with 
\begin{equation}
	\label{eq:dt_Gz}
	\Gz^{-1}_{\1 \1'} = -\vsigma_{\1 \1'}\p_{t_1} - \gamma_{\1 \1' \2}G_{\2} .
\end{equation}
Here, $g$ is to be interpreted as the \textit{bare propagator}.
In the particular case of isotropic \VRR, since ${ G_\1|_{\eta=0} \!=\! 0 }$,
we simply have ${ {\Gz^{-1}_{\1 \1'}}|_{\eta=0} \!=\! -\vsigma_{\1 \1'}\p_{t_1} }$.
Of course, premature setting of ${ \eta \!=\! 0 }$ should be avoided,
as it would result in overlooking terms coming from functional derivatives wrt $\eta$.
At this stage, it is evident that Eq.~\eqref{eq:dt_G_2_short} is not closed,
since $G_{\1 \2}$ is sourced by $G_{\1 \2 \3}$.

\subsection{Vertex renormalization}
\label{sec:Vertex}

Taking further functional derivatives of Eq.~\eqref{eq:dt_G_2_short} wrt $\eta$ leads to an unclosed statistical hierarchy, in just the same way as in Eq.~\eqref{eq:evol_C}. Indeed, the three-point cumulant, $G_{\1 \2 \3}$, explicitly depends on the three-point correlation.
In order to achieve statistical closure, one rewrites Eq.~\eqref{eq:dt_G_2_short}  in a more convenient way by expressing $G_{\1\2\3}$ in terms of $G_{\1\2}$. The key idea is to eliminate $\eta$ from the equations~\cite{MSR1973}, since $\eta$ is merely a probe of the system that perturbs the equations of motion (see Eq.~\ref{eq:dt_G_1_inter}).
To do so, one performs a change of variables from $\eta_\1$ to $G_\1$ via the Legendre transform~\citep[see section~{5.5.6} in][]{Krommes1984}
\begin{equation}
	L [G] = W[\eta] - \eta_\1 G_\1.
\label{eq:def_L}
\end{equation}
From Eq.~\eqref{eq:def_L}, one introduces the renormalized vertices as
\begin{equation}
	\Gamma_{\1 ...  \bn} = \frac{\delta^{n} L[G]}{\delta G_{\1} ... \delta G_{\bn}} ,
	\label{eq:def_Gamma}
\end{equation}
which are fully symmetric in their coordinates.
Schematically, the transformation can be summarized as
\begin{subequations}
\begin{align}
	\eta_\1 &\rightarrow G_\1\,,
	\\
	W[\eta] &\rightarrow L[G]\,,
	\\
	G_{\1...\bf{n}}  &\rightarrow \Gamma_{\1 ...  \bn} \, ,
	\label{eq:G_to_Gamma}
	\end{align}
	\label{eq:L_transform}\end{subequations}
and the vertices $\Gamma_{\1 ...  \bn}$ are now functionals of only the one-point cumulant $G_\1$. 
By taking subsequent functional derivatives of Eq.~\eqref{eq:def_L}
wrt $G_\1$, and knowing that ${ G_{\1 \2} G_{\2 \1'}^{-1} \!=\! \delta_{\1 \1'} }$,
one obtains an expression of the three-point vertex as a function of the two-point and three-point cumulants
\begin{equation}
	\Gamma_{\1 \2 \3} = G^{-1}_{\1 \1'} \,G^{-1}_{\2  \2'} \,G^{-1}_{\3 \3'} \,G_{\1' \2' \3'} .
	\label{eq:res_Gamma_3der}
\end{equation}
This is detailed in Appendix~\ref{app:Vertex_calculation}.
Importantly, the introduction of the three-point renormalized vertex  allows us to rewrite Eq.~\eqref{eq:dt_G_2_short} in terms of $\Gamma_{\1 \2 \3}$ instead of $G_{\1 \2 \3}$.
Starting from Eq.~\eqref{eq:dt_G_2_short} and using Eq.~\eqref{eq:res_Gamma_3der}, the \MSR\ closure formula finally takes the convenient form of the Dyson equation
\begin{align}
	\label{eq:MSR}
	G_{\1 \1'} =  \Gz_{\1 \1'} + \Gz_{\1 \2} \Sigma_{\2 \3} \, G_{\3 \1'}.
\end{align}
In Eq.~\eqref{eq:MSR}, we introduced the \textit{self-energy}
\begin{equation}
\Sigma_{\1 \1'} = \half \, \gamma_{\1 \2 \3} \, G_{\2 \2'} \, G_{\3 \3'} \, \Gamma_{\1' \2' \3'} ,
\label{eq:def_Sigma}
\end{equation}
and the \textit{renormalized interaction vertex}
\begin{equation}
\G_{\1 \2 \3} = \gamma_{\1 \2 \3} + \frac{\delta \Sigma_{\1 \2}}{\delta G_{\4 \5}} \, G_{\4 \4'} \, G_{\5 \5'} \, \Gamma_{\3 \4' \5'} ,
\label{eq:def_Gamma_self}
\end{equation}
as outlined in Appendix~\ref{app:Vertex_calculation}. 
With the transformation from Eq.~\eqref{eq:G_to_Gamma}, the focus of statistical closure has now shifted from determining the three-point cumulant, $G_{\1\2\3}$, to determining the three-point renormalized vertex, $\Gamma_{\1\2\3}$. In Eqs.~\eqref{eq:MSR} and~\eqref{eq:def_Gamma_self}, the replacement ${ \gamma \!\to\! \Gamma }$ (resp.\ ${ g \!\to\! G }$) is analogous to charge (resp.\ mass) renormalization in Quantum Field Theory~\citep{Peskin+2018}.
Equation~\eqref{eq:MSR} is formally exact: it is merely a sophisticated rewriting of Eq.~\eqref{eq:dt_G_2_short}.

Imposing ${ G_\1 \!=\! 0 }$ in Eq.~\eqref{eq:MSR},
and looking at the ${ ++ }$ and ${ +- }$ components of that equation, we obtain
\begin{subequations}
\begin{align}
\p_{t_1}C_{1 2} = {} & \Sigma^{\tmin \tmin}_{1 \, 3} R_{2 3} +  \Sigma^{\tmin \tplus}_{1 \, 3} C_{3 2}\, ,
\label{eq:dt_C}
\\
\p_{t_1}R_{1 2} = {} & \delta_{1 2} + \Sigma^{\tmin \tplus}_{1 \, 3} R_{3  2}\, .
\label{eq:dt_R}
\end{align}
\label{eq:dt_C_R}
\end{subequations}
This forms a set of coupled equations for the quantities of interest, that is the two-point correlation, $C$, and the mean response function, $R$.
Here, the first term in Eq.~\eqref{eq:dt_C} is a ``source'' term, and depends non-linearly on $C$ and $R$ via Eq.~\eqref{eq:def_Sigma}. In Eq.~\eqref{eq:dt_R}, this source term becomes a Dirac delta function to comply with causality. When solving Eqs.~\eqref{eq:dt_C_R}, an approximation of $\Gamma$ is required, which subsequently determines $\Sigma$ according to Eq.~\eqref{eq:def_Sigma}.

\subsection{Direct Interaction Approximation (DIA)}
\label{sec:DIA}

The \DIA\ was first introduced by~\cite{Kraichnan1958}, specifically in the context of fluid turbulence. This approach has been extended to other systems~\cite{Ottaviani1990,Krommes2002,Zhou2021}, and it is recovered by the \MSR\ formalism~\cite{MSR1973}. We refer to~\cite{Krommes2002} for a careful historical account of the \DIA\@.

As previously mentioned, one must determine the three-point vertex, $\Gamma_{\1 \2 \3}$, in order to close Eqs.~\eqref{eq:MSR}-\eqref{eq:def_Gamma_self}.
In the case of isotropic \VRR, we can consider that the characteristic \textit{charge} of our system is the bare coupling, $\gamma_{\1 \2 \3}$.
It describes how the different populations and different scales couple.
From Eqs.~\eqref{eq:def_Sigma} and~\eqref{eq:def_Gamma_self}, the simplest approximation in that parameter is
\begin{subequations}
\begin{align}
\G_{\1 \2 \3} {} & = \gamma_{\1 \2 \3} ,
\label{eq:DIA_Gamma}
\\
\Sigma_{\1 \1'} {} & = \half \, \gamma_{\1 \2 \3} \, G_{\2 \2'} \, G_{\3 \3'} \, \gamma_{\1' \2' \3'} .
\label{eq:DIA_Sigma}	
\end{align}
\label{eq:DIA_Sigma_Gamma}\end{subequations}
The general (Eulerian) \DIA\ equations are then given 
by the conjunction of the system of partial differential equations~\eqref{eq:dt_C_R}
with the approximations from Eqs.~\eqref{eq:DIA_Sigma_Gamma},
complemented with initial conditions for the functions $C$ and $R$.

The \DIA\ is the lowest-order closure scheme in the \MSR\ formalism.
Historically, when applied to isotropic fluid turbulence~\citep{Kraichnan1959},
the \DIA\ predicted an energy spectrum scaling like ${ E(k) \!\propto\! k^{-3/2} }$
for the velocity correlations,
in disagreement with the Kolmogorov's seminal prediction, ${ E (k) \!\propto\! k^{-5/3} }$~\citep{Kolmogorov1941}.
The origin for this mismatch was subsquently identified
as stemming from the fact that the original form of the \DIA\
breaks the invariance of turbulence wrt random Galilean transformations~\citep{Kraichnan1964}.
Solving this issue led to the construction of the ``Lagrangian History Direct Interaction''~\citep{Kraichnan1965},
which recovered random Galilean invariance
and subsequently Kolmogorov's spectrum.
For more detailed historical accounts,
we refer to section~{8.6.2} in~\cite{McComb1990},
section~{9.5.3} in~\cite{Frisch1995},
section~{5.6.3} in~\cite{Krommes2002}
and section~{3.2} in~\cite{Zhou2021}.

Naturally, one could be worried that these limitations of \DIA\
would impact its present application to \VRR\@.
Yet, we now point out some of the key differences
between turbulence in the Navier--Stokes system
and stochastic dynamics in \VRR\@:
(i) There is no linear term in the \VRR\ evolution equation (Eq.~\ref{eq:Master_Equation_short})
while there is for Navier--Stokes~\citep[see, e.g.\@, eq.~{(9.28)} in][]{Frisch1995}.
(ii) For \VRR\@, we consider a fully isolated system with no external perturbation.
This differs from fluid turbulence where there is, typically,
an energy injection at large scales and a subsequent cascade to small scales.
(iii) In~\cite{Kolmogorov1941}'s classical result,
one is interested in the two-point one-time correlation function
of the velocity field. For isotropic \VRR\@, this correlation is obvious (see Eq.~\ref{eq:stat_iso_C}).
Here, we are rather interested in the two-point two-time correlation
function of the system's \DF\@ (see Eq.~\ref{eq:def_C}).
(iv) Given that it involves particular orbital shapes,
\VRR\ is generically not scale-free~\citep[see appendix~{B} in][]{Kocsis+2015}.
This greatly differs from the scale-free nature of the Navier--Stokes system.
(v) Finally, given that there is no ``kinetic energy'' in \VRR\ (see Section~\ref{sec:Introduction}),
there appears to be no obvious direct analog of the random Galilean invariance
from fluid turbulence to the case of \VRR\@.
Given all these differences,
investigating whether the \DIA\ provides a good approximation
to the \VRR\ correlated dynamics appears worthwhile.

In Eq.~\eqref{eq:DIA_Sigma}, the typical amplitude of $\Sigma$ is determined by $\gamma^2$.
A (naive) higher-order expansion of $\Gamma$
would be to account for the next-order terms in powers of the bare vertex, $\gamma$,
by expanding Eq.~\eqref{eq:def_Gamma_self} wrt $\gamma$.
In Section~\ref{sec:HighOrder}, we explore this simple approach,
and show that it produces diverging predictions.
This is no surprise since \VRR\ is, in a sense, in the regime of ``strong turbulence''~\citep[see, e.g.\@][]{Krommes2002},
for which renormalizations ought to be more careful.
Self-consistent and systematic higher-order approximations~\citep[see, e.g.\@][]{MSR1973}
will be the subject of future research.

\section{Application: DIA for VRR}
\label{sec:Application}

In this section, we tailor the \DIA\ to the case of \VRR\@,
and leverage some of the specific properties of this problem.
In particular, we rely on time-stationarity and isotropy,
to considerably simplify the equations.
In addition, the strictly conservative nature of the system also plays a key role.
Finally, we compare the \DIA\ predictions with $N$-body simulations.

\subsection{Stationarity and isotropy}

Let us consider an initial distribution of orientations
that is statistically isotropic on the unit sphere,
and consists of statistically independent stellar populations.
In that case, we expect the correlation function to be stationary in time,
with an initial value set to ${ C_{{1 2\,|{ t_1\!=t_2}}} \!\!=\! \delta_{a_{1}}^{a_{2}} \deltaD(\bK_1 \!-\! \bK_2) n(\bK_1) }$.
Here, ${ n(\bK) }$ is the distribution function of the stars' parameters $\bK$
satisfying ${ \int \rd \hat{\bf{L}} \rd \bK n(\bK) \!=\! N }$~\citep[see Appendix~{D} in][]{Fouvry+2019}.
Assuming that isotropy is conserved,
the correlation can be expanded as
\begin{align}
C _{1 2} {} & = \delta_{a_1}^{a_{2}} \, \deltaD(\bK_1\!\!-\!\! \bK_2) \, C_{\ell_1} (\bK_1, t_1 \!-\! t_2) ,
\label{eq:stat_iso_C}
\end{align}
and a similar writing holds for ${ R_{1 2} \!\propto\! R_{\ell_1} }$,
with the initial conditions
${ C_{\ell}(\bK, t \!=\! 0) \!=\! n(\bK) }$
and
${ R_{\ell}(\bK, t \!=\! 0) \!=\! 1 }$. 
To further limit the complexity hidden in the coupling $\gamma_{\1\2\3}$,
we consider from now on the case of a single population system,
i.e.\ a single value of $\bK$.
In that case, one has
${ n(\bK) \!=\! N / (4 \pi) }$,
${ R_{\ell} (\bK , t) \!=\! R_{\ell} (t) }$,
${ C_{\ell} (\bK , t) \!=\!C_{\ell} (t) }$ and
${ \mJ_{\ell} [\bK , \bKp] \!=\! \mJ_{\ell} }$.
The multi-population case is briefly explored in Appendix~\ref{app:MultiPopulation}.

Our goal is now to determine a self-consistent evolution equation for the isotropic correlation function ${ t \!\mapsto\! C_{\ell} (t) }$, and compare it with numerical measurements in $N$-body simulations.

\subsection{Fluctuation-Dissipation Theorem}
\label{sec:FDT}

The isotropic \VRR\ is a thermodynamic equilibrium.
In addition, the absence of any external forcing or dissipation
makes the system strictly conservative.
For such a statistical equilibrium,
the \FDT\ holds~\citep{Kraichnan1959a},
and ensures that the correlation $C$
is directly proportional to the response function $R$.
A detailed discussion of the statistical dynamics of thermal equilibria
is provided in section~{3.7} of~\cite{Krommes2002}.
The \FDT\ for the stationary problem is also explored in~\citep{Ottaviani1990}.

The \FDT\ in the case of isotropic \VRR\ reads
\begin{align}
	\label{eq:ansatz}
	C_\ell(t) = C_0 \, R_\ell(|t|),
\end{align}
where ${ R_\ell (t) \!=\! 0 }$ for ${ t \!<\! 0 }$ and ${ C_0 \!=\! N/(4\pi) }$.
As a reassuring self-consistency check,
we explicitly verified that Eq.~\eqref{eq:ansatz}
is an exact solution of the \DIA\@, as governed by Eqs.~\eqref{eq:dt_C_R} and~\eqref{eq:DIA_Sigma_Gamma}.
In other words, for isotropic \VRR\@,
the \DIA\ is fully compatible with the \FDT\@.

As a result of the \FDT\@, Eq.~\eqref{eq:dt_R} becomes, ${ \forall \ell }$,
\begin{align}
\p_{t} R_{\ell}(t) {} & = -\frac{1}{2\ell\!+\!1}\frac{N}{4\pi} \sum_{\ell_1,\,\ell_2}[\mJ_{\ell_1} \!\!-\!\! \mJ_{\ell_2}]  [\mJ_{\ell_1}\!\! -\!\! \mJ_{\ell}] (E^L_{\ell \,\ell_1\ell_2})^2
\nonumber
\\
{} & \times \!\!\int^{t}_0\!\! \rd t' R_{\ell_1}(t \!-\! t') R_{\ell_2}(t \! - \! t') R_{\ell}(t'),
\label{eq:DIA_R_ell}
\end{align}
where the ${ \mJ_\ell \!=\! \mH_\ell/L }$ coupling coefficients and the $E^L$ Elsasser coefficients are respectively given in Appendix~\ref{App:CouplingCoefficients} and~\ref{sec:Elsasser}.
To obtain Eq.~\eqref{eq:DIA_R_ell}, we used in particular
the contraction rules of the Elsasser coefficients, as detailed in Appendix~\ref{sec:Contraction}.
This offered considerable simplifications.
Equation~\eqref{eq:DIA_R_ell} is defined for strictly positive times, ${ t \!>\! 0 }$,
with the initial condition ${ R_{\ell} (t \!=\! 0^{+}) \!=\! 1 }$.
Equation~\eqref{eq:DIA_R_ell} is the \textit{main result} of this paper.
Its generalization to a multi-population system
is provided in Appendix~\ref{app:MultiPopulation}.

Let us now comment on the main properties of Eq.~\eqref{eq:DIA_R_ell}:
(i) it is multi-scale, since different harmonics are coupled to one another;
(ii) it is highly non-linear, as the rhs scales like ${ (R_\ell)^3 }$;
(iii) it is stationary in time, as it only involves time differences;
(iv) it is non-local in time, since the integrand is evaluated at the delayed time ${ t \!-\! t' }$;
(v) it is isotropic, as all the quantities only depend on the spherical harmonic $\ell$;
(vi) because of the exclusion rules of the Elsasser coefficients and the nature of the coupling, $\mJ_\ell$, only a few terms in the double sum over harmonics are non-zero;
(vii) the number of stars $N$ can be absorbed in the definition of a new rescaled time, making the equation scale-free in $N$;
(viii) for the harmonic ${ \ell \!=\! 0 }$ (resp.\ ${ \ell \!=\! 1 }$), it yields ${ \p_{t} R_{\ell} \!=\! 0 }$, ensuring the conservation of the total mass (resp.\ total angular momentum).

\subsection{Numerical solution and $N$-body simulation}
\label{sec:Numerical}

We are now set to compare the \DIA\ predictions with numerical measurements.
Since the \VRR\ coupling coefficients typically scale like ${ \mJ_{\ellint} \!\propto\! 1/\ellint^2 }$~\citep[see appendix~{B} in][]{Kocsis+2015},
we assume, for simplicity, that only the harmonic ${ \ellint \!=\! 2 }$ contributes to the pairwise interaction.
This quadrupolar model has been shown to encompass most
of the physics of \VRR\@~\citep{Roupas+2017}.
In addition, it offers some welcome simplification of the numerics.
In Appendix~\ref{App:Quadrupolar}, we detail our numerical scheme 
to perform $N$-body simulations of the ${ \ellint \!=\! 2 }$ model
and the associated computational complexity.
We also show that our simulations are 
converged to better than ${1 \%}$.

In order to numerically solve the integro-differential Eq.~\eqref{eq:DIA_R_ell},
we proceed by (i) discretizing time regularly and evaluating the time integral
using the trapezoidal rule;
(ii) employing a second-order predictor-corrector algorithm to perform the integration per se.
Our precise scheme and computational complexity is detailed in Appendix~\ref{app:DIA_numerical}.

In Fig.~\ref{fig:DIA_Quad}, we compare the \DIA\ predictions for $R_{\ell}$~(Eq.~\ref{eq:DIA_R_ell})
with $N$-body measurements, for the ${ \ellint \!=\! 2 }$ interaction model. 
\begin{figure}[htbp!]
\begin{center}
\includegraphics[width=0.48\textwidth]{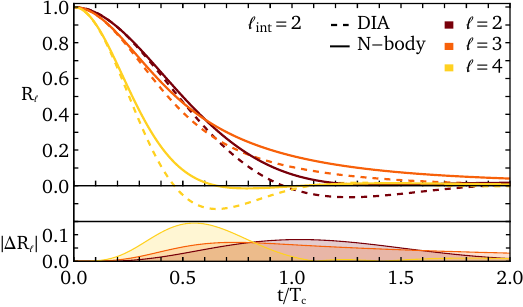}
\caption{\small{Top panel: Response function, ${ R_{\ell} }$, in the ${ \ellint \!=\! 2 }$ interaction model,
as predicted by the \DIA\ (Eq.~\ref{eq:DIA_R_ell}, dashed lines)
and measured in $N$-body simulations (full lines),
for the harmonics ${ \ell \!=\! 2,3,4 }$ (different colors). Bottom panel: Absolute error between the prediction and the measurement.
Here, the time has been rescaled by the ballistic time $\Tc$,
as defined in Eq.~\eqref{Tc}.
On short timescales,
the \DIA\ prediction agrees with the measurement,
while on longer timescales,
it overestimates the decay rate of the fluctuations.
}}
\label{fig:DIA_Quad}
	\end{center}
\end{figure}
Let us now comment on the main features of this figure.

Here, the harmonics $\ell$ correspond to the angular scales under consideration,
with larger $\ell$ corresponding to smaller angular separation between orbits.
As could have been expected, in Fig.~\ref{fig:DIA_Quad},
we find that the smaller the scale, the faster the decorrelation: large scale collective dynamics persist longer than local ones.
In that figure, we also find that, for each $\ell$ and for short times, the \DIA\ prediction
aligns with the $N$-body measurements,
with the initial derivatives matching correctly.
However, once decorrelation has started to occur,
the \DIA\ prediction seems to underestimate the numerical measurements.
In addition, for even harmonics, ${ \ell \!=\! 2, 4 }$,
the predictions change sign.
This is reminiscent of the behavior already observed in fig.~{3}
of~\cite{Kraichnan1961}, where the \DIA\ was applied for one of the first times. It is worth noting that for ${\ell\!=\!4}$,
 the $N$-body measurement becomes slightly negative at ${t/\Tc \!\sim\! 0.65}$,
 indicating an anticorrelation.
Given the numerous (and strong) assumptions of the \DIA\ low-order closure scheme,
Fig.~\ref{fig:DIA_Quad} offers overall a satisfactory result.

In Appendix~\ref{app:multi_ell_int},
we briefly explore the behavior of the \DIA\ prediction
should one account for all the harmonics ${ \ellint \!\geq\! 2 }$.
The associated result is presented in Fig.~\ref{fig:DIA_Asymp},
where we find that a sharper interaction potential
leads to faster decorrelations.

\subsection{Higher-order approximation?}
\label{sec:HighOrder}

In Section~\ref{sec:DIA}, we focused on the \DIA\@,
the lowest-order truncation in the \MSR\ formalism.
We now briefly discuss the difficulties arising
when considering possible higher-order approximations of Eq.~\eqref{eq:DIA_R_ell}.

\begin{figure}[htbp!]
\begin{center}
\includegraphics[width=0.48\textwidth]{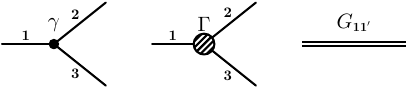}
\\
\includegraphics[width=0.48\textwidth]{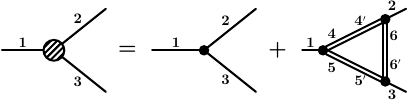}
\caption{Diagrammatic representation of the \MSR\ formalism.
Top: the bare vertex, $\gamma_{\1\2\3}$ (left), the renormalized vertex, $\Gamma_{\1\2\3}$ (middle), and the propagator, $G_{\1 \1'}$ (right).
Bottom: graphical representation of the (naive) next-order expansion from Eq.~\eqref{eq:Gamma_next_order}.
This expansion is shown to misbehave in Appendix~\ref{App:Higher_order}.}
\label{fig:Diagrams}
\end{center}
\end{figure}

Following Eq.~\eqref{eq:DIA_Sigma_Gamma},
for the \DIA\@, we simply have ${ \Gamma \!=\! \gamma }$ and ${ \Sigma \!\propto\! \gamma^2 }$.
A naive iteration would be to consider the next-order contribution in $\gamma^4$
within the expression of $\Sigma$ from Eq.~\eqref{eq:def_Sigma},
i.e.\ to perform an expansion of $\Gamma$ wrt $\gamma$.
This expansion is depicted in fig.~{2.(b)} of~\cite{MSR1973},
and is also explored in the context of path-integral formalisms in eq.~{(68)} of~\cite{Valageas2004}.
This approach is particularly simplistic because it assumes that $\gamma$
can be treated as a perturbative parameter and that the expansion actually converges
-- which can be the case in regimes of ``weak turbulence''~\citep[see, e.g.\@][]{Krommes2002}.
Using this approach, the next-order expansion of $\Gamma$ would read
\begin{equation}
	\label{eq:Gamma_next_order}
	\Gamma_{\1 \2 \3} = \gamma_{\1 \2 \3} + \gamma_{\1 \4 \5} G_{\4 \4'} \gamma_{\2 \4' \6} G_{\5 \5'} \, G_{\6 \6'} \, \gamma_{\3 \5' \6'} .
\end{equation}
A diagrammatic representation of Eq.~\eqref{eq:Gamma_next_order}
is presented in Fig.~\ref{fig:Diagrams}. 

\begin{figure}[htbp!]
\includegraphics[width=0.48\textwidth]{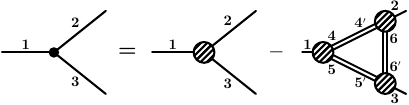}
\caption{Diagrammatic representation of the improved next-order expansion of the bare vertex, $\gamma_{\1\2\3}$, in powers of the renormalized vertex, $\Gamma_{\1\2\3}$.}
\label{fig:Diagrams_improved}
\end{figure}

When plugged into Eq.~\eqref{eq:def_Sigma},
the attempt from Eq.~\eqref{eq:Gamma_next_order}
would ultimately lead to an expansion of $\Sigma$ in terms of $\gamma^2$ and $\gamma^4$,
as detailed in Eq.~\eqref{eq:Sigma_next_order}.
Pushing further the calculation, one can write explicitly
the next-order integro-differential equation satisfied by $R_{\ell}$,
as given in Eq.~\eqref{eq:dt_R_next_order}.
When solved numerically, the associated prediction shows
signs of strong divergence,
as illustrated in Fig.~\ref{fig:MSR_Quad}. This behavior is strikingly similar to that of fig.~{5} in~\cite{Kraichnan1961}.
Such a divergence was (somewhat) expected
given the inherent naivety of the expansion from Eq.~\eqref{eq:Gamma_next_order}.
Indeed, it assumes that $\gamma$ can be treated as a meaningful
perturbative parameter: this is surely no given in the present
regime of isotropic \VRR\@. 
As a conclusion, in order to improve upon the \DIA\ prediction
from Fig.~\ref{fig:DIA_Quad}, a more careful and self-consistent
next-order renormalization scheme
must be implemented to offer a converging prediction.
This is discussed in~\cite{MSR1973,Krommes2002},
which argue that, in the regime of ``strong turbulence'',
one should rather expand $\gamma$ in terms of $\Gamma$ (see Fig.~\ref{fig:Diagrams_improved}).
Such an improved approach then requires
to solve simultaneously joint equations for $\Gamma$ and $G$.
This will be the focus of future work.

\subsection{Previous works}
\label{sec:PreviousWorks}

It is interesting to compare the present approach
to previous methods from the literature
on \VRR\@.

A first detailed investigation
of the statistics of \VRR\ was made in~\cite{Kocsis+2015},
in particular in section~{4} therein.
They proceed by introducing two parameters:
(i) a ballistic time, $t_{\rms}$,
namely the time it would take for the test star to traverse the entire sphere
should it feel a constant (and typical) torque~\citep[see eq.~{71} in][]{Kocsis+2015};
(ii) a decoherence time, $t_{\phi}$,
so that on time intervals shorter than $t_{\phi}$,
the torque felt by the test star is temporally correlated,
while this torque decorrelates on timescales longer than $t_{\phi}$~\cite[see eq.~{78} in][]{Kocsis+2015}.
Importantly, for isotropic \VRR\@, these two timescales
follow the exact same scaling wrt to the total number
of particles $N$.
\VRR\ is then modeled 
as a time-correlated random walk on the unit sphere.
In particular, \cite{Kocsis+2015} recovered
both the ballistic and diffusive regimes of diffusion,
as illustrated in figs.~{10}--{12} therein.
Even if based on motivated heuristics,
the approach from~\cite{Kocsis+2015}
recovered all the important qualitative features
of the numerical measurements.

The statistics of isotropic \VRR\ was also explored in~\cite{Fouvry+2019},
through the following approximations.
First, one computes
the initial time-derivative, ${ \p_{t}^{2} C_{\ell} / \p t^{2} [t \!=\! 0] \!\propto\! 1 / T_{\ell} }$,
assuming Gaussian statistics.
Second, one approximates the time-dependence of the correlation as a Gaussian,
i.e.\ ${ C_{\ell}^{\rG} (t) \!\propto\! \re^{- (t / T_{\ell})^{2} / 2}  }$~\citep[see eq.~{23} in][]{Fouvry+2019}.
Third, one considers the problem of a zero-mass test particle
embedded within a Gaussian random noise following
the statistics of $C_{\ell}^{\rG}$.
The correlation of the test particle's evolution,
${ C_{\ell}^{\test} (t) \!\propto\! \langle \varphi_{\ell}^{\test} (0) \varphi^{\test}_{\ell} (t) \rangle }$,
is then approximated as ${ C_{\ell}^{\test} \!=\! C_{\ell}^{\test} [C_{\ell}^{\rG}] }$~\citep[see eq.~{33} in][]{Fouvry+2019}.
Fourth, one introduces back self-consistency,
by assuming that the dynamics of each (massive) particle
can be approximated by the one of a zero-mass test particle.
This leads to a self-consistent relation of the form
${ C_{\ell} \!=\! C_{\ell} [C^{\test}] }$~\citep[see eq.~{44} in][]{Fouvry+2019}.
\cite{Fouvry+2019} compared their predictions
with $N$-body measurements,
and could recover the exponential tail of the correlation function,
see fig.~{3} therein.

\begin{figure}[htbp!]
\begin{center}
\includegraphics[width=0.48\textwidth]{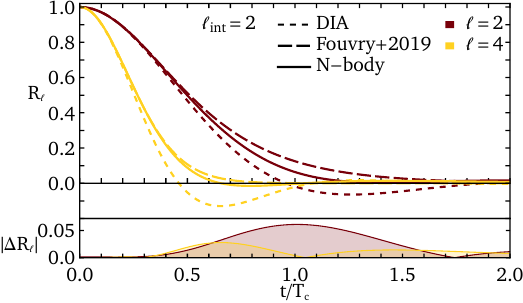}
\caption{\small{Top panel: Same as Fig.~\ref{fig:DIA_Quad},
with the predictions from~\citep[][eq.~{44} therein, large dashes]{Fouvry+2019}. Bottom panel: Absolute error between the prediction from~\cite{Fouvry+2019} and the $N$-body measurement.
In practice, \cite{Fouvry+2019} tends to overestimate the $N$-body, while the \DIA\ underestimates it.
For scale ${\ell\!=\!4}$, \cite{Fouvry+2019} is a better fit but misses the negative part of the correlation. 
}}
\label{fig:DIA_vs_Fouvry}
	\end{center}
\end{figure}

In Fig.~\ref{fig:DIA_vs_Fouvry}, we compare the \DIA\ predictions
with eq.~{(44)} from~\cite{Fouvry+2019}.
Importantly, \cite{Fouvry+2019} considered solely even harmonics, so only the scales ${\ell\!=\!2,4}$ are represented here.
For ${\ell\!=\!2}$, the prediction from~\cite{Fouvry+2019} overestimates the measurement,
while the \DIA\ underestimates it.
For ${\ell\!=\!4}$, \cite{Fouvry+2019} provides a better fit to the $N$-body compared to the \DIA\@.
Yet, \cite{Fouvry+2019} fails to capture the negative part of the correlation,
which the \DIA\ does reproduce. 
One of the virtues of the present \MSR\ approach is that systematic higher-order approximations
seem within (reasonable) reach. In contrast, it is far from obvious
how the approach from~\cite{Fouvry+2019}
may be iterated upon to lead ever better approximations.

\section{Conclusion}
\label{sec:Conclusion}

In this paper, we investigated the correlated dynamics of \VRR\@,
as an archetype of quadratically non-linear stochastic dynamics.
In particular, we emphasized how the generic \MSR\ formalism,
and its leading order limit, the \DIA\@, can be explicitly implemented
to estimate the two-point correlation function of fluctuations.
We applied this approach in the isotropic limit of \VRR\@,
for which the \FDT\ holds -- a welcome simplification.
Our main result was obtained in Eq.~\eqref{eq:DIA_R_ell},
a fully explicit integro-differential equation for the two-point correlation functions
of the fluctuations.
This was complemented with Fig.~\ref{fig:DIA_Quad},
where we compared this prediction with direct measurements
in numerical simulations. We found the two approaches to be
in satisfactory agreement, though the \DIA\ prediction
overestimated the decay rate of the correlation functions.

The present work is only a first step toward a thorough
application of statistical closure theory to the problem of \VRR\@.
Let us conclude by mentioning a few possible venues
for future works.

It surely would be worthwhile to develop the present
scheme to the next order. This requires some ``vertex renormalization'',
i.e.\ the self-consistent determination of $\Gamma_{\1\2\3}$~\citep[see fig.~{3} in][]{MSR1973},
contrary to the \DIA\ which sets it to the bare vertex (see Eq.~\ref{eq:DIA_Gamma}).
While more challenging, we expect for this calculation
to remain (reasonably) tractable. For instance, we expect the \FDT\ to hold at any order,
as a direct consequence of the thermodynamic equilibrium that is isotropic \VRR\@.

Naturally, a legitimate question is to wonder whether this approach
can be further iterated upon, ad libitum.
Yet, this is unfortunately not given for the \MSR\@ formalism.
Indeed, in the absence of any small parameter
controlling clearly the order of the expansion,
iterations on the \MSR\ scheme
may, in a similar fashion to Section~\ref{sec:HighOrder},
diverge at higher order~\citep[see, e.g.\@, fig.~{5} in][]{Kraichnan1961}.
In that context, one could investigate the possibility
of more intricate higher-order renormalizations
using $n$-body functions and the associated
Bethe--Salpeter equations~(see, e.g.\@, eq.~{234} in~\citep{Krommes1984}
and section~{3.1} in~\citep{Fukuda1995}),
and $n$-particule irreducible effective actions~\citep[see, e.g.\@,][]{Cornwall1974,Berges2004,Carrington2011}.

Equation~\eqref{eq:Master_Equation_short},
the fundamental evolution equation for \VRR\@,
drives some form of cascade between large and small scales, as does Eq.~\eqref{eq:DIA_R_ell}.
For example, as noted in Fig.~\ref{fig:DIA_Quad},
large $\ell$, i.e.\ smaller scales, decorrelate faster.
In addition, given that the \VRR\ interaction is dominated by the (large-scale)
quadrupolar ${ \ellint \!=\! 2 }$ interaction,
the dynamics of some given harmonics, $\ell$,
is dominated by the interaction of the (squeezed) triplets of scales
${ \{\ell, \ell_{1} , \ell_{2} \} \!\sim\! \{ \ell , 2 , \ell \!\pm\! 1 \} }$.
In that context, it would be interesting to investigate
the applicability of ``renormalization group'' methods~\citep[see, e.g.\@,][for a review]{Zhou2010}
to \VRR\@.

Similarly, one should also investigate the use
of methods stemming
from the ``non-perturbative functional renormalization group''~\citep[see, e.g.\@,][for a review]{Delamotte2012,Dupuis+2021}.
In this approach, contributions from ever larger scales are progressively
accounted for by adding a carefully chosen ``regulator''
to the system's action.
In addition, ``functional renormalization'' places a particular emphasis
on leveraging the system's (extended) symmetries~\citep[see, e.g.\@,][]{Canet2022,Fontaine+2023}.
Interestingly, \cite{Tarpin+2019} used such an approach to predict
the two-point correlation function in the related context
of stationary and isotropic two-dimensional turbulence.
On that front, the similarities between eq.~{(59)} of~\cite{Tarpin+2019}
and the \VRR\ result given in eq.~{(33)} of~\cite{Fouvry+2019}
are particularly striking.

Finally, in the astrophysical context, it would be interesting
to consider more realistic setups, by alleviating some of our
simplifying assumptions. In particular, in no specific order,
one should:
(i) go beyond the assumption of a statistically
isotropic distribution of orientations~\citep[see, e.g\@,][]{Roupas+2017,Touma+2019,Gruzinov+2020};
(ii) consider the impact of different stellar populations,
e.g.\@, different masses,
possibly leading to the spontaneous formation of stellar discs~\citep[see, e.g.\@,][]{Szolgyen+2018,Magnan+2022,Mathe+2023};
(iii) investigate the role played by different harmonics
contributing to the gravitational coupling~\citep{Takacs+2018};
(iv) account for the central \BH\@'s
relativistic Lense--Thirring precession~\citep[see, e.g.\@,][]{Fragione+2022};
(v) characterize the impact of a possible intermediate mass \BH\
also orbiting within the system~\citep{Gravity+2023,Will+2023};
(vi) investigate the relaxation of a single (zero-mass) test star
embedded in this fluctuating system~\citep{Kocsis+2015,Fouvry+2019};
(vii) examine the importance of the dynamical friction
undergone by a massive test particle in this environment~\citep{Ginat+2022};
(viii) describe the respective diffusion of two test stars
away from one another, i.e.\ ``neighbor separation''~\citep{Giral+2020}.
Overall, the goal of these various developments would be to use
the observation of the (clockwise) stellar disc around SgrA*~\citep[see, e.g.\@,][]{Paumard+2006,Bartko+2009,Lu+2009,Yelda+2014,Gillessen+2017,vonFellenberg+2022}
along with a precise characterization of \VRR\
to constrain the content of SgrA*'s stellar cluster~\citep[see, e.g.\@,][]{Panamarev+2022,Fouvry+2023}.

\begin{acknowledgments}
This work is partially supported by the grant Segal ANR-19-CE31-0017 
of the French Agence Nationale de la Recherche
and by the Idex Sorbonne Universit\'e.
This research was supported in part by grant NSF PHY-2309135
to the Kavli Institute for Theoretical Physics (KITP).
We warmly thank B.~Deme, A.~El Rhirhayi,
J.~Magorrian, C.~Pichon, M.~Roule, A.~Schekochihin
for stimulating discussions. We thank the two anonymous referees for their insightful suggestions that greatly improved the paper.
\end{acknowledgments}

\appendix

\section{Vector Resonant Relaxation}
\label{App:VRR}

\subsection{Coupling coefficients}
\label{App:CouplingCoefficients}

Following~\cite{Magnan+2022} and references therein,
the coupling coefficients appearing in Eq.~\eqref{rewrite_Htot} read
\begin{equation}
\mH_{\ell} [\bK , \bKp] \!=\! \frac{4 \pi G m m'}{2 \ell \!\!+\!\! 1} |P_{\ell} (0)|^{2} \!\! \int_{0}^{\pi} \!\!\frac{\rd M}{\pi} \frac{\rd M'}{\pi} \!\frac{\Min [r , r']^{\ell}}{\Max [r , r']^{\ell + 1}} ,
\label{exp_mH}
\end{equation}
with $P_{\ell}$ the Legendre polynomial of order $\ell$
and ${ (M , M') }$ the mean anomalies of the orbits.
We note that
(i) all odd harmonics $\ell$ are associated with ${ \mH_{\ell} \!=\! 0 }$;
(ii) the harmonic ${ \ell \!=\! 0 }$ does not drive any dynamics;
(iii) the coefficients satisfy the symmetry
${ \mH_{\ell} [\bK , \bKp] \!=\! \mH_{\ell} [\bKp , \bK] }$.

To simplify the writing,
it is convenient to also introduce the rescaled
coupling coefficients
\begin{equation}
\mJ_{\ell} [\bK , \bKp] = \mH_{\ell} [\bK , \bKp] / L (\bK),
\label{exp_mJ}
\end{equation}
used later in Eq.~\eqref{eq:def_Gamma_alpha_0}.

\subsection{ Equations of motions}
\label{app:Equations_motion}

Following~\cite{Kocsis+2015}, the dynamics of a given (zero-mass)
test particle is given by the effective Hamiltonian
\begin{align}
	\label{eq:H_single}
	\Ht = - \sum_{i=1}^{N} \sum_{a} \mH_{a} [\bK , \bK_{i}] \, Y_{a} (\hbL) \, Y_{a} (\hbL_{i}),
\end{align}
with ${ a \!=\! (\ell,m) }$.

Let us now introduce the usual ${ (\theta, \phi) }$ spherical coordinates. Because of the double orbit-average from Eq.~\eqref{eq:H_single},
we may track the evolution with the canonical coordinates,
${ (\phi, L_z \!=\! L\cos \theta) }$. As a result, for \VRR\@, phase space is the unit sphere.
The individual equations of motion are given by Hamilton's equations
\begin{equation}
	\frac{\rd \phi}{\rd t} = \frac{\p \Ht}{\p L_z}; \quad \frac{\rd L_z}{\rd t} = -\frac{\p \Ht}{\p \phi}.
	\label{eqs_motion_ham}
\end{equation}
Introducing the operator
\begin{equation}
	\label{eq:gradient}
	\hbL \times  \frac{\p}{\p \hbL} = \bf{e}_r \!\times\! \left\{ \frac{\p}{\p \theta} \bf{e}_\theta + \frac{1}{\sin{\theta}}\frac{\p}{\p \phi}\bf{e}_\phi\right\} ,
\end{equation}
within the basis ${ (\bf{e}_r,\bf{e}_\theta,\bf{e}_\phi) }$,
we get from Eq.~\eqref{eqs_motion_ham} 
\begin{equation}
\frac{\rd \hbL}{\rd t} = - \frac{1}{L [\bK]} \, \hbL \!\times\! \frac{\p \Ht}{\p \hbL} .
\label{EOM_ind}
\end{equation}
This equation of motion is the one appearing in Eq.~\eqref{evol_vphid}. 

\subsection{Elsasser coefficients}
\label{sec:Elsasser}

\subsubsection{Definition}
Following appendix~{B} in~\cite{Fouvry+2019},
the Elsasser coefficients are defined with the convention
\begin{equation}
E_{a b c} = \!\! \int \!\! \rd \hbL \, Y_{a} (\hbL) \, \bX_{b} (\hbL) \!\cdot\! \frac{\p Y_{c} (\hbL)}{\p \hbL},
\label{def_Elsasser}
\end{equation}
with ${ a \!=\! (\ell,m) }$ and the (real) vector spherical harmonics,
${ \bX_{a} \!=\! \hbL \!\times\! \p Y_{a} (\hbL) / \p \hbL }$, with ${ \p / \p \hbL }$ given in Eq.~\eqref{eq:gradient}.
These coefficients are generically decomposed in~\citep{Fouvry+2019}
\begin{equation}
E_{abc} = E^L_{\ell_{a} \ell_{b} \ell_{c}} \, E^{M}_{m_a m_b m_c} .
\label{decomp_Elsasser}
\end{equation}
The $E^M$ (resp.\ $E^L$) coefficients are antisymmetric (resp.\ symmetric) when any two indices are transposed. Remarkably, the coefficients $E_{a b c}$
are zero unless all pairs ${ (\ell,m) }$ are different and
\begin{equation}
|\ell_{a} \!-\! \ell_{b}| < \ell_{c} < \ell_{a} \!+\! \ell_{b}
\;\; \text{and} \;\;
\ell_{a} \!+\! \ell_{b} \!+\! \ell_{c}
\; \text{is odd} .
\label{triangular_test}
\end{equation}

\subsubsection{Contraction rules}
\label{sec:Contraction}

We follow Chap.~{12} of~\cite{Varshalovich1988}
to obtain appropriate contraction rules
for the anisotropic Elsasser coefficients.

To obtain Eq.~\eqref{eq:DIA_R_ell},
we use the identity
\begin{equation}
\sum_{\mathclap{m_{c} , m_{d}}} E^{M}_{a c d} \, E^{M}_{b c d} =  \frac{\delta_{a}^{b}}{2 \ell_{a} + 1} ,
\label{contract_2}
\end{equation}
where, importantly, it is assumed that all the interaction pairs
appearing in Eq.~\eqref{contract_2},
i.e.\ ${ (a,c,d) }$ and ${ (b, c, d) }$,
satisfy the criteria from Eq.~\eqref{triangular_test}.

To derive the (naive) higher-order approximation
from Eq.~\eqref{eq:dt_R_next_order},
we use the relation
\begin{equation}
\sum_{\mathclap{\substack{m_{c} , m_{d} \\ m_{e} , m_f \\ m_g}}} E^{M}_{a c d} \, E^{M}_{b e f}  \, E^{M}_{g c f} \, E^{M}_{g e d} =  \frac{\delta_{a}^{b}}{2 \ell_{a} \!+\! 1} W^{\ell_a \, \ell_{c} \, \ell_{d}}_{\ell_g \, \ell_{e} \, \ell_{f}},
\label{contract_4}
\end{equation}
where we introduced Wigner's ${ 6j }$ symbols
\begin{align}
\label{eq:Wigner}
W^{\ell_a \, \ell_{c} \, \ell_{d}}_{\ell_g \, \ell_{e} \, \ell_{f}} = (- 1)^{\ell_{c} + \ell_{e}}
\scalebox{0.9}{$\begin{Bmatrix}
\ell_{a} & \ell_{c} & \ell_{d}
\\
\ell_{g} & \ell_{e} & \ell_{f}
\end{Bmatrix}$}.
\end{align}
Here again, all the triplets of interaction appearing
in Eq.~\eqref{contract_4} must satisfy the conditions
from Eq.~\eqref{triangular_test}.

\subsection{Bare interaction coefficient}
\label{app:Interaction_coeff}

The non-random bare interaction coefficient appearing in Eq.~\eqref{eq:Master_Equation}
is generically given by
\begin{align}
\gamma_{a b c} [\bK_{a}, \bK_{b},\bK_{c}]  = E_{a b c} \, \big\{ {} & \mJ_{\ell_c} [\bK_{a} , \bK_{c}] \deltaD \big[ \bK_{a} \!\! -\!\! \bK_{b} \big]
\nonumber
\\
- {} & \mJ_{\ell_b} [\bK_{a}, \bK_{b}] \deltaD \big[ \bK_{a} \!\!-\!\! \bK_{c} \big] \big\} ,
\label{eq:def_Gamma_alpha_0}
\end{align}
with $\mJ_\ell$ provided by Eq.~\eqref{exp_mJ}.

Including the time coordinate as ${ 1 \!=\! (a, \bK_{a}, t_{a}) }$, the generalized bare interation coefficient from Eq.~\eqref{eq:Master_Equation_short} is then
\begin{equation}
	\gamma_{1 2 3} = \gamma_{a b c} [\bK_{a}, \bK_{b},\bK_{c}]\, \deltaD ( t_{a} \!-\! t_{b} ) \, \deltaD ( t_{a} \!-\! t_{c} ) .
	\label{eq:def_Gamma_0}
\end{equation}
It is symmetric in its last to arguments, i.e.\ ${ \gamma_{1 2 3} \!=\! \gamma_{1 3 2} }$.

\section{Martin-Siggia-Rose Closure}
\label{app:MSR}

\subsection{Bare interaction vertex}
\label{app:Interaction_vertex}

In Eq.~\eqref{eq:evol_Phi}, we introduce the generalized coordinate ${ \1 \!=\! (\eps, 1) }$
with ${ \eps \!=\! \pm }$.
The bare interaction vertex reads
\begin{align}
\gamma_{\1 \2 \3} = {} & \half \big\{ \gamma_{1 2 3} \!+\! \gamma_{1 3 2}  \big\} \, \delta_{\eps_{1}}^{-} \, \delta_{\eps_{2}}^{+} \delta_{\eps_{3}}^{+}
\nonumber
\\
+ {} & \half \big\{ \gamma_{2 1 3} \!+\! \gamma_{2 3 1} \big\} \, \delta_{\eps_{2}}^{-} \, \delta_{\eps_{1}}^{+} \, \delta_{\eps_{3}}^{+}
\nonumber
\\
+ {} & \half \big\{ \gamma_{3 2 1} \!+\! \gamma_{3 1 2} \big\} \, \delta_{\eps_{3}}^{-} \, \delta_{\eps_{2}}^{+} \, \delta_{\eps_{1}}^{+} .
\label{eq:def_Gamma_generic}
\end{align}
Importantly, $\gamma_{\1\2\3}$ is fully symmetric in its arguments,
i.e.\ it is left unchanged by any permutation of two indices.

\subsection{Response function}
\label{app:Response}

In this appendix, we closely follow section 5.5.6 of~\cite{Krommes1984}. To precisely define the system's response to an infinitesimal fluctuation $\eta$, we rewrite the equation of motion Eq.~\eqref{eq:Master_Equation_short} for the distribution function $\vphi$, while adding an external source $\eta$.
This reads
\begin{equation}
\label{eq:Master_pert}
\p_{t} \vphi_1 = \half \, \gamma_{1 2 3} \, \vphi_2 \, \vphi_3 + \eta_1,
\end{equation}
where $\eta_1$ corresponds to $\eta^{\tmin}_1$ defined in Eq.~\eqref{eq:def_Z}.
Let us now define the stochastic response function, describing the response of the system at point $1$ to an infinitesimal fluctuation at point $1'$.
It reads
\begin{equation}
	\label{eq:def_R_tilde}
	\tR_{1 1'} = \delta \vphi_1 / \delta \eta_{1'} \big|_{\eta=0} \quad \forall \, t_1 > t_{1'} .
\end{equation}
By differentiating Eq.~\eqref{eq:Master_pert} wrt $\eta_{1'}$, we obtain 
\begin{equation}
\label{eq:R_tilde}
\p_{t} \tR_{1 1'} = \half \, \gamma_{1 2 3} \, ( \vphi_2 \tR_{3 1'} + \tR_{2 1'} \vphi_3) + \delta_{1 1'}.
\end{equation}
Although the definition in Eq.~\eqref{eq:def_R_tilde} is explicit, it is not practical since the quantity of real interest is the mean response function ${ R \!=\! \ml \tR \mr }$. Averaging Eq.~\eqref{eq:R_tilde} would result in intricate terms like ${ \ml \vphi \tR \mr }$. Therefore, an alternative representation of the response function is preferred. Specifically, one that gives $R$ the same structure as $C$,
for these two quantities to be conjugate.

In the \MSR\ formalism, the key idea is to introduce the operator $\hvphi$ through the canonical commutation relation
\begin{equation}
	\label{eq:commutation}
	[\vphi(\u{1},t); \hvphi(\u{1}',t)] = \delta_{\u{1} \, \u{1}'},
\end{equation}
where the time coordinate is excluded by defining ${ 1 \!=\! (\u{1},t) }$ and ${ [A;B] \!=\! AB \!-\! BA }$.
This relation is analogous to ${ [q ; p]\!=\! \ri \hbar }$ in Quantum Mechanics,
with $q$ the position and ${ p \!=\! - \ri \hbar \p_q }$ the momentum~\citep[see, e.g.\@,][]{Binney+2013}.
An equivalent path-integral formulation of \MSR\ is reviewed in section~{6.4} of~\cite{Krommes2002}.

We further introduce 
\begin{equation}
	\tr_{1 1'} = \Theta(t-t') [\vphi_1 ; \hvphi_{1'}],
\end{equation}
with $\Theta$ the usual Heaviside function.
Using Eqs.~\eqref{eq:Master_Equation_short} and~\eqref{eq:commutation}
along with the identity ${ [AB;C] \!=\! A[B;C] \!+\! [A;C]B }$,
we get
\begin{equation}
\label{eq:r_tilde}
\p_{t} \tr_{1 1'} = \half \, \gamma_{1 2 3} \, ( \vphi_2 \tr_{3 1'} + \tr_{2 1'} \vphi_3) + \delta_{1 1'}.
\end{equation}
Even though $\tR$ and $\tr$ satisfy the same evolution equation, we cannot simply state that ${ \tR \!=\! \tr }$.
Indeed, the operator $\tr$ does not commute with $\vphi$,
whereas $\tR$ does.
However, as argued in~\cite{Krommes1984}, these two quantities are equal when ensemble-averaged, i.e.\ ${ r \!=\! \ml \tr \mr \!=\! R }$, provided (i) an appropriate definition of the operator $\hvphi$ and (ii) the condition ${ \ml \hvphi \, ... \mr \!=\! 0 }$ is satisfied, where ``${...}$'' is any combination of $\vphi$ and $\hvphi$ .
The mean response function is then given by
\begin{equation}
	R_{1 1'} = \Theta(t-t') \ml [\vphi_1 ; \hvphi_{1'}] \mr = \Theta(t-t') \ml \vphi_1 \,\hvphi_{1'} \mr.
\end{equation}
From ${ \ml \hvphi \mr \!=\! 0 }$, we can write
\begin{subequations}
\begin{align}
	\label{eq:Response_conjugate}
	R_{1 1'} {} & = \ml \vphi_1 \,\hvphi_{1'} \mr - \ml \vphi_1 \mr \ml \hvphi_{1'} \mr, \quad \forall \; t > t',
\\
	\label{eq:C_conjugate}
	C_{1 1'} {} & = \ml \vphi_1 \,\vphi_{1'} \mr - \ml \vphi_1 \mr \ml \vphi_{1'} \mr, \quad \forall \; t\,,t'.
\end{align}
\label{eq:RC_conjugate}\end{subequations}
We thus have a clear symmetric writing between the two-point correlation, $C$, and the mean response function, $R$. Equations~\eqref{eq:Response_conjugate} and~\eqref{eq:C_conjugate} are completely equivalent to the expressions given in Eq.~\eqref{eq:C_R_W}.

\subsection{Field equations}
\label{app:field_equations}

The commutation relation for the generalized two-component field $\Phi$
follows from Eq.~\eqref{eq:commutation}.
It reads
\begin{equation}
	\label{eq:def_Phi_commutator}
	[\Phi(\eps_1,\u{1},t); \Phi(\eps_2,\u{2},t)] = \vsigma_{\eps_1 \eps_2} \delta_{\u{1} \, \u{2}},
\end{equation}
with the Pauli-like tensor, $\vsigma$, appearing in Eq.~\eqref{eq:evol_Phi},\begin{align}
	{} & \vsigma_{\tplus \tplus} = 0 ; \quad \quad \vsigma_{\tplus \tmin} = 1 ; 
	\nonumber
	\\
	{} & \vsigma_{\tmin \tplus} = - 1 ; \,  \quad \vsigma_{\tmin \tmin} = 0 .
	\label{eq:def_vsigma}
\end{align}

\subsection{Cumulant dynamics}
\label{app:mean_field}

In this appendix, we explicitly derive evolution equations for the one-point  and two-point cumulants $G_\1$ and $G_{\1 \2}$.
We follow the approach from~\cite{Krommes2002}.
From Eq.~\eqref{eq:def_cumulants}, we have
\begin{equation}
\label{eq:def_G_1}
G_{\1} = \frac{\delta W[\eta]}{\delta \eta_{\1}} = \frac{\ml \Phi_\1 \exp{(\Phi_\2 \eta_\2)} \mr }{\ml\exp{(\Phi_\2 \eta_\2)}\mr}.
 \end{equation}
Separating the time coordinate like ${ \1 \!=\! (\u{\1},t) }$,
and dealing carefully with time-ordering (from right to left~\cite{MSR1973}), we can write 
\begin{align}
	\p_{t} \ml \Phi(\u{\1},t) \exp{(\Phi_\2 \eta_\2)} \mr {} &= \p_{t} \ml \exp{[\scalebox{0.95}{$\int^{\infty}_{t}$} \rd t' \Phi(\u{\2},t')\eta(\u{\2},t')]} 
	\nonumber
	\\
	{} & \!\!\!\!\!\!\!\! \times \Phi(\u{\1},t) \exp{[\scalebox{0.95}{$\int^{t}_{-\infty}$} \rd t' \Phi(\u{\2},t')\eta(\u{\2},t')]}\mr
	\nonumber
	\\
	{} &  = \ml \dot{\Phi}(\u{\1},t) \exp{[\Phi_\2 \, \eta_\2]} \mr 
	\nonumber
	\\
	{} & + \ml [-\Phi(\u{\2},t)\eta(\u{\2},t)\Phi(\u{\1},t) 
	\label{eq:calc_EOM_G1}
	\\
	{} & + \Phi(\u{\1},t)\Phi(\u{\2},t)\eta(\u{\2},t)] \exp{[\Phi_\2 \, \eta_\2]}\mr.
	\nonumber
\end{align}
Using the commutation relation from Eq.~\eqref{eq:def_Phi_commutator}, we get
\begin{align}
	\label{eq:dt_G_1_inter}
	\p_{t} G(\u{\1},t) {} &= \frac{\ml \dot{\Phi}(\u{\1},t) \exp(\Phi_\2 \eta_\2 ) \mr}{\ml \exp{(\Phi_\2 \eta_\2) \mr}} + \vsigma \eta(\u{\1},t),
\end{align}
where ${\vsigma \eta(\u{\1},t) \!=\! \vsigma_{\eps_1 \eps_2} \eta(\eps_2,\u{1},t)}$.
We now note that
\begin{equation}
	G_\3G_\4 + G_{\3 \4} = \frac{\ml \Phi_\3 \Phi_\4 \exp{(\Phi_\5 \eta_\5)} \mr}{\ml \exp{(\Phi_\5 \eta_\5)} \mr},
\end{equation}
and inject Eq.~\eqref{eq:evol_Phi} into Eq.~\eqref{eq:dt_G_1_inter}
to obtain
\begin{equation}
	\label{eq:dt_G_1}
	\p_{t} G_\1 = \half \vsigma_{\1 \2}\gamma_{\2 \3 \4} \big( G_\3G_\4 + G_{\3 \4} \big) + \vsigma \eta_\1.
\end{equation}
Here, one should pay attention to the source term ${ \vsigma \eta_{\1} }$
stemming from time-ordering.
Since ${ G_{\1 \1'} \!=\! \delta G_\1 / \delta \eta_{\1'} }$ (see Eq.~\ref{eq:def_cumulants}),
differentiating Eq.~\eqref{eq:dt_G_1} wrt $\eta_{\1'}$ yields
\begin{equation}
	\label{eq:dt_G_2_app}
	\p_{t_1} G_{\1 \1'} = \vsigma_{\1 \2} \gamma_{\2 \3 \4} G_{\3} G_{\4 \1'} + \half \vsigma_{\1 \2} \gamma_{\2 \3 \4} G_{\3 \4 \1'} + \vsigma_{\1 \1'}.
\end{equation}
Introducing the identity tensor ${ \delta_{\1 \1'} \!=\! -\vsigma_{\1 \2}\vsigma_{\2 \1'} }$,
we can rewrite Eq.~\eqref{eq:dt_G_2_app} as 
\begin{equation}
	\label{eq:dt_G_2_short_app}
	G_{\1 \1'} =  \Gz_{\1 \1'} + \half \Gz_{\1 \2} \gamma_{\2 \3 \4} G_{\3 \4 \1'},
\end{equation}
with 
\begin{equation}
	\label{eq:dt_Gz_app}
	\Gz^{-1}_{\1 \1'} = -\vsigma_{\1 \1'}\p_{t_1} - \gamma_{\1 \1' \2}G_{\2} .
\end{equation}

\subsection{Vertex calculation}
\label{app:Vertex_calculation}

We compute in this appendix the first vertices,
as introduced in Eq.~\eqref{eq:def_Gamma}.
To do so, we take subsequent functional derivatives wrt $G_\1$ of the Legendre transform $L$ defined in Eq.~\eqref{eq:def_L}.
Considering that the quantity ${ W[\eta] }$ only depends on $G_\1$ through the mapping ${ \eta_\1 \!\rightarrow\! G_\1 }$, we can write ${ \delta/\delta G_\1 \!=\! \delta/\delta \eta_\2 \big|_{G_\1} \! \delta \eta_\2/ \delta G_\1 }$. It follows
\begin{subequations}
\begin{align}
\Gamma_\1 {} & = \frac{\delta W[\eta]}{\delta \eta_\2}\frac{\delta \eta_\2}{\delta G_\1} -\frac{\delta \eta_\2}{\delta G_\1} G_\2 -\eta_\1  = - \eta_\1,
\label{eq:res_Gamma_1der}
\\
\Gamma_{\1 \2} {} & = - \frac{\delta \eta_\2}{\delta G_\1} = - G^{-1}_{\1 \2}.
\label{eq:res_Gamma_2der}
\end{align}
\label{eq:res_Gamma}\end{subequations}

Let us now calculate the three-point vertex $\Gamma_{\1\2\3}$,
as given by Eq.~\eqref{eq:res_Gamma_3der}.
First, we differentiate Eq.~\eqref{eq:res_Gamma_2der} wrt ${ G_\3 }$
to write
\begin{equation}
\Gamma_{\1 \2 \3} = - \frac{\delta G^{-1}_{\1 \2}}{\delta G_\3} . 
\label{calc_Gamma_3der}
\end{equation}
To make progress, we use the identity operator
\begin{equation}
	\label{eq:identity}
	G_{\1 \2} \,G^{-1}_{\2  \1'} = \delta_{\1 \1'}.
\end{equation}
Multiplying Eq.~\eqref{eq:identity} on the left by $G$ and differentiating it wrt $\eta$ yields
\begin{equation}
	G_{\1 \1'} \, G_{\2 \2'}\frac{\delta G^{-1}_{\1 \2}}{\delta \eta_{\3'}} = -G_{\1' \2' \3'}.
\end{equation}
Then, noting that
\begin{equation}
	\frac{\delta G^{-1}_{\1 \2}}{\delta G_{\3}} = \frac{\delta G^{-1}_{\1 \2}}{\delta \eta_{\3'}} \frac{\delta \eta_{\3'}}{\delta G_{\3}},
\end{equation}
since $G^{-1}_{\1 \2}$ depends on $G_\3$ through $\eta$, we ultimately obtain
\begin{equation}
	\Gamma_{\1 \2 \3} = G^{-1}_{\1 \1'} \, G^{-1}_{\2 \2'} \, G^{-1}_{\3 \3'} \, G_{\1' \2' \3'} .
\end{equation}

We now want to derive the expression of the three-point vertex given in Eq.~\eqref{eq:def_Gamma_self}. To do so, we start from the Dyson equation~\eqref{eq:MSR}
\begin{equation}
	G^{-1}_{\1 \2} = \Gz^{-1}_{\1 \2} - \Sigma_{\1 \2},
\end{equation}
and differentiate it wrt $G_\3$, thus giving
\begin{equation}
	\label{eq:exp_Gamma}
	\Gamma_{\1 \2 \3} = \gamma_{\1 \2 \3} + \frac{\delta \Sigma_{\1 \2}}{\delta G_\3}.
\end{equation}
In Eq.~\eqref{eq:exp_Gamma}, we used the definition from Eq.~\eqref{eq:dt_Gz}.
Assuming from Eqs.~\eqref{eq:exp_Gamma} and~\eqref{eq:def_Sigma} that $\Sigma_{\1 \2}$ depends on $G_\3$ only explicitly through $G_{\4 \5}$ \cite[see section 6.2.3 in][]{Krommes2002}, we may write
\begin{equation}
	\label{eq:diff_Sigma}
	\frac{\delta \Sigma_{\1 \2}}{\delta G_\3}\bigg|_{\eta} \!\!=\!\frac{\delta \Sigma_{\1 \2}}{\delta G_{\4 \5}}\bigg|_{G_\3} \! \! \frac{\delta G_{\4 \5}}{\delta G_{\3}}\bigg|_\eta \!\! = \!\frac{\delta \Sigma_{\1 \2}}{\delta G_{\4 \5}}G_{\4 \4'} G_{\5 \5'} \Gamma_{\3 \4' \5'},
\end{equation}
where we used the identity from Eq.~\eqref{eq:identity}.
Plugging Eq.~\eqref{eq:diff_Sigma} into Eq.~\eqref{eq:exp_Gamma} yields Eq.~\eqref{eq:def_Gamma_self}.

\section{Multi-population and DIA}
\label{app:MultiPopulation}

In the case of a multi-population system,
the \FDT\ from Eq.~\eqref{eq:ansatz} becomes
\begin{align}
	\label{eq:ansatz_multi}
	C_{\ell}(\bf{K},t) = n(\bK) R_\ell(\bK, |t|),
\end{align}
with ${ n(\bK) }$ the distribution function of the stars' parameters,
${ \bK \!=\! (m,a,e) }$.

After some manipulations, the equivalent of Eq.~\eqref{eq:DIA_R_ell} for a multi-population system is
\begin{align}
	{} & \p_t R_{\ell}(\bf{K},t) 
	\nonumber
	\\
	{} & =\! -\frac{1}{2 \ell + 1}\!\!\sum_{\ell_1,\ell_2}\!\!\int_0^{t}\!\!\rd t'\!\!\int\!\! \rd \bf{K}' \!\bigg[n(\bf{K}')\mJ_{\ell_1}[\bf{K},\bf{K}']R_{\ell}(\bf{K},t')
	\nonumber
	\\
	{} & - n(\bf{K})\mJ_{\ell_2}[\bf{K}',\bf{K}]R_{\ell}(\bf{K}',t')\bigg] \mJ_{\ell_1}[\bf{K},\bf{K}'](E^L_{\ell \, \ell_1 \ell_2 })^2 
	\nonumber
	\\
	{} & \times R_{\ell_1}(\bf{K}',t\!\!-\!\!t')R_{\ell_2}(\bf{K},t\!\!-\!\!t').
	\label{eq:DIA_R_ell_multi}
\end{align}
Such a multi-population expression
should prove particularly useful to model the Galactic Center,
where a wide range of different stellar populations are present~\citep{Ghez+2008,Gillessen+2017}.

\section{The single-population ${ \ellint \!=\! 2 }$ model}
\label{App:Quadrupolar}
We provide here more detail on the ${ \ellint \!=\! 2 }$ interaction model,
introduced in Section~\ref{sec:Numerical}. 
We consider \VRR\ in the limit
of a single population,
interacting only via the ${ \ellint \!=\! 2 }$ harmonics~\citep[see, e.g.\@,][]{Roupas+2017}.
Assuming that all orbits are identical and circular,
they share the same parameter
\begin{equation}
\bK_{\star} = (m_{\star} , a_{\star} , e_{\star} \!=\! 0) ,
\label{def_Kstar}
\end{equation}
so that ${ M_{\star} \!=\! N m_{\star} }$ is the total stellar mass.
Following Eq.~\eqref{exp_mH},
the ${ \ell \!=\! 2 }$ coupling coefficient reads
\begin{align}
\mH_{2} {} & = \mH_{2} [\bK_{\star} , \bK_{\star}] = \frac{\pi}{5} \, \frac{G m_{\star}^{2}}{a_{\star}} ,
\label{exp_H2}
\end{align}
and the individual norm of the angular momentum vectors is
${ |\bL_{i}| \!=\! L_{0} \!=\! m_{\star} \sqrt{G M_\bullet a_{\star}} }$.

\subsection{Equations of motion}
\label{sec:EOM_quadrupole}

Using the addition theorem for spherical harmonics,
the Hamiltonian from Eq.~\eqref{rewrite_Htot} generically becomes~\cite{Kocsis+2015}
\begin{equation}
\Htot = - \sum_{i < j}^{N} \sum_{\ell} \frac{2 \ell + 1}{4 \pi} \mH_{\ell} [\bK_{i} , \bK_{j}] \, P_{\ell} (\hbL_{i} \!\cdot\! \hbL_{j}) .
\label{rewrite_Htot_Legendre}
\end{equation}
Following Eq.~\eqref{EOM_ind}, the individual equations of motion
then read
\begin{equation}
\frac{\rd \hbL_{i}}{\rd t} = \! \sum_{j = 1}^{N} \sum_{\ell \geq 2} \! \frac{2 \ell + 1}{4 \pi L[\bK_{i}]} \mH_{\ell} [\bK_{i} , \bK_{j}] P_{\ell}^{\prime} (\hbL_{i} \!\cdot\! \hbL_{j}) \hbL_{i} \!\times\! \hbL_{j} .
\label{EOM_Legendre}
\end{equation}
In the particular case of the ${ \ellint \!=\! 2 }$ model,
one has ${ P_{2} (x) \!=\! \tfrac{1}{2} (3 x^{2} \!-\! 1) }$,
and Eq.~\eqref{EOM_Legendre} gives
\begin{equation}
\frac{\rd \hbL_{i}}{\rd t} = \frac{1}{L_{0}} \frac{15}{4 \pi} \, \mH_{2} \, \sum_{j = 1}^{N} \big( \hbL_{i} \!\cdot\! \hbL_{j} \big) \, \hbL_{i} \!\times\! \hbL_{j} .
\label{EOM_Legendre_next}
\end{equation}

Let us now define the (symmetric) matrix
\begin{equation}
\bM = \sum_{i = 1}^{N} \hbL_{i} \otimes \hbL_{i} .
\label{def_bM}
\end{equation}
We emphasize that $\bM$ can be computed in ${ \mathcal{O} (N) }$ operations.
This makes $N$-body explorations of the quadrupolar ${ \ellint \!=\! 2 }$ model quite inexpensive.
We can write
\begin{align}
\big[ \bM \, \hbL_{i} \big]_{p} \!\!=\!\! \sum_{q} \! \sum_{j} \!\!\big[ \hbL_{j} \big]_{p} \, \!\!\big[ \hbL_{j} \big]_{q} \, \!\!\big[ \hbL_{i} \big]_{q}\!\! =\!\! \bigg[ \!\!\sum_{j} \big( \!\hbL_{i} \!\cdot\! \hbL_{j} \!\big) \hbL_{j} \bigg]_{p}\!\!.
\label{contraction_M}
\end{align}
Equation~\eqref{EOM_Legendre_next} finally becomes
\begin{equation}
\frac{\rd \hbL_{i}}{\rd t} = \omega_{0} \, \hbL_{i} \!\times\! \big( \bM \, \hbL_{i} \big) ,
\label{EOM_Legendre_final}
\end{equation}
with the frequency scale 
\begin{equation}
\omega_{0} = \frac{3}{4} \sqrt{\frac{G M_\bullet}{a_{\star}^{3}}} \, \frac{m_{\star}}{M_\bullet} .
\label{def_Omega_0}
\end{equation}
Finally, following eq.~{(19)} of~\cite{Fouvry+2019}, for that model,
the typical decay rate of the correlation is the ballistic time
\begin{equation}
	\Tc = \frac{4 \sqrt{5}}{\sqrt{3}} \, \sqrt{\frac{a_{\star}^{3}}{G \MBH}} \, \frac{\MBH}{M_{\star}} \sqrt{N} .
	\label{Tc}
\end{equation}

\subsection{Time integration}
\label{sec:TimeIntegration}
Given that the \VRR\ dynamics conserves ${ |\hbL_{i}| }$ for every particle,
we can rewrite the evolution equations as
\begin{equation}
\frac{\rd \hbL}{\rd t} = \bO (\hbL) \!\times\! \hbL ,
\label{rewrite_EOM_nbody}
\end{equation}
with the (conservative) choice, ${ \bO \!=\! \hbL \!\times\! \rd \hbL / \rd t }$.
To integrate the system forward in time,
we then use a structure-preserving integration scheme similar
to the one presented in section~{5} of~\cite{Fouvry+2022}.
This is now briefly detailed.

For a fixed value of $\bO$ and an initial condition $\hbL_{0}$,
Eq.~\eqref{rewrite_EOM_nbody} can be integrated exactly for a duration $t$
to the new location
\begin{align}
\hbL (t) {} & = \cos (\Omega t) \hbL_{0} \!\!+\! \sin (\Omega t) \, \hbO \!\times\! \hbL_{0} \!\!+\! [1 \!-\! \cos (\Omega t) ] (\hbL_{0} \!\cdot\! \hbO) \hbO
\nonumber
\\
{} & = \phi [t \bO] \circ \hbL_{0} ,
\label{Rodrigues}
\end{align}
with ${ \Omega \!=\! |\bO| }$ and ${ \hbO \!=\! \bO / \Omega }$.
Equation~\eqref{Rodrigues} ensures that ${ |\bL (t)| \!=\! |\bL_{0}| }$.
In practice, we apply this formula for ${ |\bL_{0}| \!=\! 1 }$,
and systematically perform the renormalization
${ \hbL \!\leftarrow\! \hbL / |\hbL| }$
after every evaluation of Eq.~\eqref{Rodrigues}.
This prevents a drift of ${ |\hbL| }$ through round-off errors.

\begin{figure}[htbp!]
\begin{center}
\includegraphics[width=0.48\textwidth]{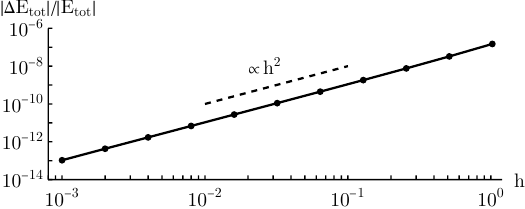}
\caption{\small{Relative error in the total energy ${ \Etot (t\!=\!10.24) }$
in one $N$-body simulation,
using the same convention as in Fig.~\ref{fig:DIA_Quad},
as a function of the timestep $h$.
Relative errors are computed wrt ${ \Etot (t \!=\! 0) }$.
As expected, the integration scheme from Eq.~\eqref{eq:def_MD2}
is second-order accurate.
\label{fig:Convergence_Nbody}}}
\end{center}
\end{figure}

Now that we have this generic ``drift'' operator at our disposal,
we use a simple two-stage explicit midpoint rule.
Given some timestep $h$,
and starting from some initial stage $\hbL_{n}$,
it proceeds via
\begin{subequations}
\begin{align}
\bO_{1} {} & = \bO (\hbL_{n}) ,
\label{eq:bO1}
\\
\hbL_{2} {} & = \phi [\tfrac{h}{2} \bO_{1}] \circ \hbL_{n} ,
\label{eq:hL2}
\\
\bO_{2} {} & = \bO (\hbL_{2}) ,
\label{eq:bO2}
\\
\hbL_{n+1} {} & = \phi [h \bO_{2}] \circ \hbL_{n} .
\label{eq:Ln+1}
\end{align}
\label{eq:def_MD2}\end{subequations}
The scheme from Eq.~\eqref{eq:def_MD2}
is (i) explicit;
(ii) conserves ${ |\hbL| \!=\! 1 }$ exactly;
(iii) requires two computations of the rates of change;
(iv) is second-order accurate, see~\citep{Fouvry+2022}.
This is checked in Fig.~\ref{fig:Convergence_Nbody}. 

In practice, to obtain the $N$-body measurement in Fig.~\ref{fig:DIA_Quad},
we used ${ G \!=\! M_{\bullet} \!=\! M_{\star} \!=\! a_{\star} \!=\! 1 }$,
and ${ N \!=\! 1\,000 }$.
The numerical integration was performed using
${ h \!=\! 10^{-3} }$, with a dump every ${ \Delta t \!=\! 1 }$
up to ${ t_{\max} \!=\! 10^{5} }$.
With these choices, the final relative error in the total energy
(resp.\ total angular momentum) was of the order
$10^{-12}$ (resp.\ $10^{-10}$).
On a single core, one such simulation required ${\sim 3.5}$h.
We used a total of ${10\,000}$ independent realizations
to ensemble average.
Performing a bootstrap resampling over the available realizations
and comparing with the expected value for zero-time separation,
we could check that the correlation functions
measured in $N$-body simulations were converged
to better than ${ 1\% }$.

Once the dynamics of the particles has been integrated,
it remains to compute the harmonic coefficients, ${ \vphi_{a} (t) }$
with ${ a \!=\! (\ell , m) }$.
Following Eq.~\eqref{expansion_vphid}, they read
\begin{equation}
\vphi_{a} (t) = \sum_{i = 1}^{N} Y_{a} \big[ \hbL_{i} (t) \big] .
\label{eq:calc_phi_a}
\end{equation}
In practice, to compute the (real) spherical harmonics,
we use the exact same recurrences as in appendix~{C} of~\cite{Fouvry+2019}.

At this stage, for each realization and each harmonics ${ a \!=\! (\ell , m) }$,
we have at our disposal a time series of the form
${ \{ \vphi_{a} (n \Delta t) \}_{0 \leq n \leq \nmax} }$.
Owing to time stationarity,
the auto-correlation of each of these time series
is estimated via
\begin{equation}
C_{a} [t \!=\! n \Delta t] \!=\! \tfrac{1}{\nmax - n + 1} \!\!\! \sum_{i=0}^{\nmax - n} \!\!\! \vphi_{a} [i \Delta t] \vphi_{a} [(i \!+\! n) \Delta t] .
\label{eq:calc_correl}
\end{equation}
Finally, for every $\ell$, we average
over $m$ and over realizations. This is the $N$-body
result presented in Fig.~\ref{fig:DIA_Quad}.

\section{Numerical integration of DIA}
\label{app:DIA_numerical}

\begin{figure}[htbp!]
\begin{center}
\includegraphics[width=0.48\textwidth]{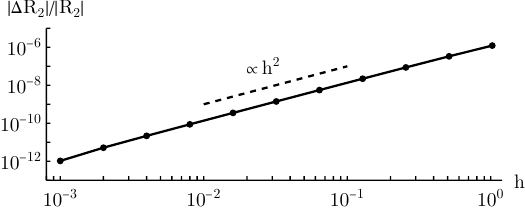}
\caption{\small{Relative error in ${ R_{2} (t \!=\! 10.24) }$,
using the same convention as in Fig.~\ref{fig:DIA_Quad}
(with ${ \ellmax \!=\! 10 }$),
as a function of the timestep $h$.
Relative errors are computed wrt the calculation with ${ h \!=\! 5 \!\times\! 10^{-4} }$.
As expected, the integration scheme from Eqs.~\eqref{trapezoidal_rule}
and~\eqref{calc_f} is second-order accurate.
\label{fig:Convergence_DIA}}}
\end{center}
\end{figure}

In this appendix, we detail the integration scheme implemented to solve Eq.~\eqref{eq:DIA_R_ell}.
It is a generic non-linear integro-differential equation
of the form
\begin{equation}
\frac{\p R}{\p \tau} = \!\! \int_{0}^{\tau} \!\! \rd s \, W [s, \tau] ;
\quad R(\tau \!=\! 0) = 1 .
\label{shape_eq}
\end{equation}
where $W[s,\tau]$ depends on the values of ${ R (s) }$ within the time interval ${ 0 \!\leq\! s \!\leq\! \tau }$.
Let us now give ourselves a certain timestep $h$, and discretize time as ${ \tau_{i} \!=\! i \, h }$
with ${ i \!\geq\! 0 }$.
The integral in the rhs of Eq.~\eqref{shape_eq}
is evaluated using the trapezoidal rule. This reads
\begin{align}
\int_{0}^{\tau_{n}} \!\! \rd s \, W [s , \tau_{n}] {} & \simeq h \bigg\{ \sum_{i = 1}^{n-1} W [\tau_{n-i} , \tau_{n}]
\nonumber
\\
{} & \quad\quad \!+\! \tfrac{1}{2} W [\tau_{0},\tau_{n}] \!+\! \tfrac{1}{2} W [\tau_{n} , \tau_{n}] \bigg\}
\nonumber
\\
{} & = I_{n} \!\big[ R_{0}, ... , R_{n} \big] ,
\label{trapezoidal_rule}
\end{align}
with the shortened notation ${ R_{i} \!=\! R (\tau_{i}) }$.
Note that for ${ n \!=\! 0 }$, Eq.~\eqref{trapezoidal_rule}
is (slightly) corrected to read ${ I_{0} [R_{0}] \!=\! 0 }$.
With this approximation, we now use a second-order
predictor-corrector algorithm to integrate Eq.~\eqref{shape_eq}.
We first compute the prediction
\begin{equation}
\hR_{n+1} = R_{n} + h \, I_{n} \!\big[ R_{0}, ... , R_{n} \big] ,
\label{calc_hf}
\end{equation}
which is then corrected via
\begin{equation}
R_{n+1} \!=\! R_{n} \!+\! \tfrac{h}{2} \big( I_{n} \!\big[ R_{0}, ... , R_{n} \big] \!+\! I_{n+1} \!\big[ R_{0}, ... , R_{n} , \hR_{n+1} \big] \big) .
\label{calc_f}
\end{equation}
This scheme is second-order,
i.e.\ the error after some finite time scales like ${ \mO (h^{2}) }$.
This is checked in Fig.~\ref{fig:Convergence_DIA}.

In practice, the double summation over spherical harmonics appearing in Eq.~\eqref{eq:DIA_R_ell} is (abruptly) truncated
by setting ${ R_{\ell} \!=\! 0 }$ for all ${ \ell \!>\! \ellmax }$.
We checked that for low harmonics $\ell$,
the prediction remains unchanged
when considering a sufficiently high $\ellmax$.

For a given $\ellmax$ and a given number of timesteps $N_{t}$,
the overall complexity of this algorithm scales like ${ \mO (\ellmax^{3} N_{t}^{2}) }$.
Fortunately, the computation can be somewhat accelerated
by carefully accounting for the various exclusion rules
of the isotropic Elsasser coefficients $E^L$, see Eq.~\eqref{decomp_Elsasser}. 
The associated numerical code is publicly distributed~\cite{github}.

To obtain the prediction presented in Fig.~\ref{fig:DIA_Quad}, we fixed units by setting ${ G \!=\! M_{\bullet} \!=\! M_{\star} \!=\! a_{\star} \!=\! 1 }$,
${N\!=\!1\,000}$, in which case the ballistic time is ${ \Tc \!\simeq\! 163 }$,
see Appendix~\ref{App:Quadrupolar}.
We also set ${ h \!=\! 1.0}$, ${ \ellmax \!=\! 20}$. 
We checked that varying $h$ and $\ellmax$ (${10 \!\leq\! \ellmax \!\leq\! 100}$)
did not affect the predicted correlations for ${2\!\leq\!\ell\!\leq\!5}$ in the ${\ellint\!=\!2}$ model.
Specifically, with these parameters,
the computation time for the \DIA\ scheme
is ${\leq \! 1 \mathrm{s}}$ on 8 threads.

\section{DIA and the ${ \ellint \!\geq\! 2 }$ interaction model}
\label{app:multi_ell_int}

\begin{figure}[htbp!]
\begin{center}
\includegraphics[width=0.48\textwidth]{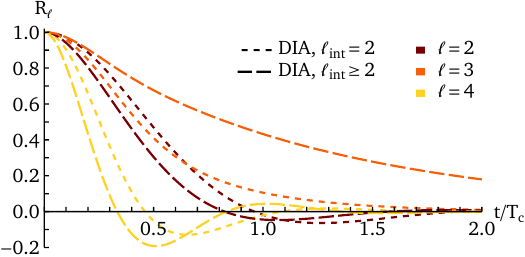}
\caption{\small{Same as Fig.~\ref{fig:DIA_Quad},
except that we compare the \DIA\ prediction (Eq.~\ref{eq:DIA_R_ell})
for the ${ \ellint \!=\! 2 }$ model (small dashes)
with the ${ \ellint \!\geq\! 2 }$ model (large dashes).
For even harmonics, the correlations decay faster 
as one includes more harmonics in the pairwise interaction.
\label{fig:DIA_Asymp}}}
\end{center}
\end{figure}

In this appendix, we briefly explore the single-population model
when all harmonics ${ \ellint \!\ge\! 2 }$ are taken into account.
Following appendix~{B} in~\cite{Kocsis+2015},
we assume, for simplicity, that the coupling coefficients scale like
${ \mJ_{\ellint} \!=\! \mJ_{2} / ([\ellint / 2])^{2} }$.
In that case, we follow the same approach as in Appendix~\ref{app:DIA_numerical}
to compute the \DIA\ prediction,
with the added complexity of many more terms in the double harmonics sums.
In Fig.~\ref{fig:DIA_Asymp}, we compare the predictions obtained through \DIA\ for both models. 

In that figure, the predictions for both models exhibit the same overall behavior,
though they differ in their respective correlation times.
Indeed, for even harmonics, the model ${ \ellint \!\geq\! 2 }$ decorrelates faster.
Following eq.~{(15)} of~\cite{Fouvry+2019},
we argue that this is because the typical correlation time of a given (even) harmonic
is inversely proportional to the sum ${ \sum_{\ellint}\!\!\mJ^2_{\ellint} }$~\citep[see][]{Fouvry+2019}. As a result, the sharper the interaction potential,
the faster the decorrelation for the even harmonics.
These even harmonics are the sole driver of the dynamics of particles,
see after Eq.~\eqref{exp_mH}.

\section{Higher-order approximation?}
\label{App:Higher_order}

\begin{figure}[htbp!]
\begin{center}
\includegraphics[width=0.48\textwidth]{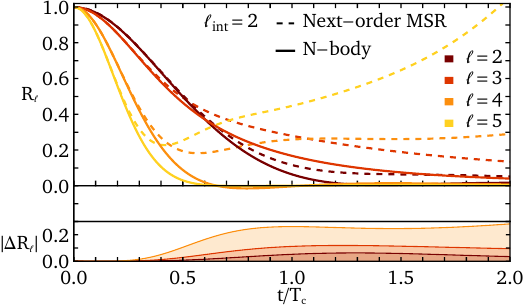}
\caption{\small{Same as Fig.~\ref{fig:DIA_Quad},
but for the (naive) higher-order prediction (Eq.~\ref{eq:dt_R_next_order}, dashed lines). 
On short timescales, the higher-order prediction
agrees with the measurements.
However, the prediction diverges on longer timescales,
and the divergence worsens as the scale becomes smaller.
}}
\label{fig:MSR_Quad}
	\end{center}
\end{figure}

Here, we provide the next-order analytical and numerical resolution of Eq.~\eqref{eq:Gamma_next_order}.
Injecting Eq.~\eqref{eq:Gamma_next_order} into the definition
from Eq.~\eqref{eq:def_Sigma}, we obtain
\begin{align}
	\label{eq:Sigma_next_order}
	\Sigma_{\1 \1'} {} & = \half \, \gamma_{\1 \2 \3} \, G_{\2 \2'} \, G_{\3 \3'} \, \gamma_{\1' \2' \3'} 
	\\
	\nonumber
	{} & +  \half \, \gamma_{\1 \2 \3} \, G_{\2 \2'} \, G_{\3 \3'} \, \gamma_{\1' \4 \5} G_{\4 \4'} \gamma_{\2' \4' \6} G_{\5 \5'} \, G_{\6 \6'} \, \gamma_{\3' \5' \6'}.
\end{align}
We now inject Eq.~\eqref{eq:Sigma_next_order} into Eqs.~\eqref{eq:dt_C} and~\eqref{eq:dt_R}, and assume stationarity and isotropy as in Eq.~\eqref{eq:stat_iso_C}.
In that case, we recover that the \FDT\ also holds:
this is once again a reassuring self-consistency check.
After lengthy manipulations,
the prediction for the isotropic response function, $R_\ell$,
in a single population system becomes
\begin{widetext}
\begin{align}
	\label{eq:dt_R_next_order}
	\frac{\p \mR_{\ell_1}(t_0)}{\p t_0} = {} & - \frac{1}{2\ell_1 + 1} C_0 \sum_{\ell_2, \ell_3} [\mJ_{\ell_2} \!\!-\!\! \mJ_{\ell_3}]  [\mJ_{\ell_2} \!\!-\!\! \mJ_{\ell_1}] (E^L_{\ell_1\ell_2\ell_3})^2\!\!\int^{t_0}_0\!\! \rd t \, \mR_{\ell_1}(t) \mR_{\ell_2}(t_0\!\!-\!\! t) \mR_{\ell_3}(t_0\!\!-\!\! t)  
	\\
	\nonumber
	{} & +  \frac{1}{2\ell_1 + 1} C_0^2 \sum_{\mathclap{\substack{\ell_2 , \ell_3 \\ \ell_4 , \ell_5, \ell_6 }}}\, [\mJ_{\ell_1} \!\!-\!\! \mJ_{\ell_5}][\mJ_{\ell_2} \!\!-\!\! \mJ_{\ell_3}]\,E^{L}_{\ell_1\ell_2\ell_3} \, E^{L}_{\ell_1\ell_5\ell_6} \, E^{L}_{\ell_4\ell_2\ell_6} \, E^{L}_{\ell_4\ell_5\ell_3} \, W^{\ell_1 \, \ell_2 \, \ell_3}_{\ell_4 \,\ell_5 \,\ell_6}
	\\
	\nonumber
	{} & \times \!\!\int^{t_0}_0\!\! \rd t_1 \!\!\int^{t_0}_0\!\! \rd t_2 \!\!\int^{t_0}_0\!\! \rd t_3 \,  \mR_{\ell_1}(t_3)\,\mR_{\ell_2}(t_0 \!\!-\!\! t_2) \, \mR_{\ell_3}(t_0 \!\!-\!\! t_1) \, \mR_{\ell_4}(t_1\!\! -\!\! t_2) \,  \mR_{\ell_5}(t_1 \!\!-\!\! t_3) \, \mR_{\ell_6}(t_2 \!\!-\!\! t_3) \,
	\\
	\nonumber
	{} & \times \!\!\bigg\{ \!\![\mJ_{\ell_5} \!\!-\!\! \mJ_{\ell_4}][\mJ_{\ell_4} \!\!-\!\! \mJ_{\ell_6}]\,\Theta(t_0\!\!-\!\!t_1)\Theta(t_0\!\!-\!\!t_2)\Theta(t_2\!\!-\!\!t_3) + 	[\mJ_{\ell_4} \!\!-\!\! \mJ_{\ell_3}][\mJ_{\ell_4} \!\!-\!\! \mJ_{\ell_6}]\,\Theta(t_0\!\!-\!\!t_2)\Theta(t_2\!\!-\!\!t_3)\Theta(t_3\!\!-\!\!t_1)
	\\
	\nonumber
	{} & + [\mJ_{\ell_5} \!\!-\!\! \mJ_{\ell_4}][\mJ_{\ell_6} \!\!-\!\! \mJ_{\ell_2}]\,\Theta(t_0\!\!-\!\!t_1)\Theta(t_1\!\!-\!\!t_2)\Theta(t_2\!\!-\!\!t_3) + [\mJ_{\ell_4} \!\!-\!\! \mJ_{\ell_6}][\mJ_{\ell_3} \!\!-\!\! \mJ_{\ell_5}]\,\Theta(t_0\!\!-\!\!t_2)\Theta(t_2\!\!-\!\!t_3)\Theta(t_2\!\!-\!\!t_1)\!\!\bigg\}.
\end{align}
\end{widetext}
The first term on the rhs is exactly Eq.~\eqref{eq:DIA_R_ell}. Here, $\Theta$ is the usual Heaviside function and the coefficients $W^{\ell_1 \, \ell_2 \, \ell_3}_{\ell_4 \, \ell_5 \, \ell_6}$ are given by Eq.~\eqref{eq:Wigner}. The numerical solution of Eq.~\eqref{eq:dt_R_next_order} is illustrated in Fig.~\ref{fig:MSR_Quad}, in the case of the quadrupolar ${ \ellint \!=\! 2 }$ interaction model. 
In Fig.~\ref{fig:MSR_Quad}, we find that the higher-order prediction
from Eq.~\eqref{eq:dt_R_next_order} severely diverges at late times.
And, that this divergence worsens as one considers larger $\ell$,
i.e.\ smaller angular scales.

As discussed in Section~\ref{sec:HighOrder}, the divergence observed
in Fig.~\ref{fig:MSR_Quad}
originates from the (incorrect) assumption that the expansion
of the dressed vertex, $\Gamma$,
in terms of the bare vertex, $\gamma$, converges.
A brief discussion on a more promising approach to derive the higher-order prediction is mentioned in Section~\ref{sec:HighOrder}.
This will be the focus of future work.


\begin{thebibliography}{78}
\expandafter\ifx\csname natexlab\endcsname\relax\def\natexlab#1{#1}\fi
\expandafter\ifx\csname bibnamefont\endcsname\relax
  \def\bibnamefont#1{#1}\fi
\expandafter\ifx\csname bibfnamefont\endcsname\relax
  \def\bibfnamefont#1{#1}\fi
\expandafter\ifx\csname citenamefont\endcsname\relax
  \def\citenamefont#1{#1}\fi
\expandafter\ifx\csname url\endcsname\relax
  \def\url#1{\texttt{#1}}\fi
\expandafter\ifx\csname urlprefix\endcsname\relax\def\urlprefix{URL }\fi
\providecommand{\bibinfo}[2]{#2}
\providecommand{\eprint}[2][]{\url{#2}}

\bibitem[{\citenamefont{{Kormendy} and {Ho}}(2013)}]{Kormendy+2013}
\bibinfo{author}{\bibfnamefont{J.}~\bibnamefont{{Kormendy}}} \bibnamefont{and}
  \bibinfo{author}{\bibfnamefont{L.~C.} \bibnamefont{{Ho}}},
  \bibinfo{journal}{ARA\&A} \textbf{\bibinfo{volume}{51}}, \bibinfo{pages}{511}
  (\bibinfo{year}{2013}).

\bibitem[{\citenamefont{{Ghez} et~al.}(2008)}]{Ghez+2008}
\bibinfo{author}{\bibfnamefont{A.~M.} \bibnamefont{{Ghez}}}
  \bibnamefont{et~al.}, \bibinfo{journal}{ApJ} \textbf{\bibinfo{volume}{689}},
  \bibinfo{pages}{1044} (\bibinfo{year}{2008}).

\bibitem[{\citenamefont{{Gillessen} et~al.}(2017)}]{Gillessen+2017}
\bibinfo{author}{\bibfnamefont{S.}~\bibnamefont{{Gillessen}}}
  \bibnamefont{et~al.}, \bibinfo{journal}{ApJ} \textbf{\bibinfo{volume}{837}},
  \bibinfo{eid}{30} (\bibinfo{year}{2017}).

\bibitem[{\citenamefont{{Paumard} et~al.}(2006)}]{Paumard+2006}
\bibinfo{author}{\bibfnamefont{T.}~\bibnamefont{{Paumard}}}
  \bibnamefont{et~al.}, \bibinfo{journal}{ApJ} \textbf{\bibinfo{volume}{643}},
  \bibinfo{pages}{1011} (\bibinfo{year}{2006}).

\bibitem[{\citenamefont{{Murchikova} et~al.}(2019)\citenamefont{{Murchikova},
  {Phinney}, {Pancoast}, and {Blandford}}}]{Murchikova+2019}
\bibinfo{author}{\bibfnamefont{E.~M.} \bibnamefont{{Murchikova}}},
  \bibinfo{author}{\bibfnamefont{E.~S.} \bibnamefont{{Phinney}}},
  \bibinfo{author}{\bibfnamefont{A.}~\bibnamefont{{Pancoast}}},
  \bibnamefont{and} \bibinfo{author}{\bibfnamefont{R.~D.}
  \bibnamefont{{Blandford}}}, \bibinfo{journal}{Nature}
  \textbf{\bibinfo{volume}{570}}, \bibinfo{pages}{83} (\bibinfo{year}{2019}).

\bibitem[{\citenamefont{{GRAVITY Collaboration} et~al.}(2020)}]{Gravity+2020}
\bibinfo{author}{\bibnamefont{{GRAVITY Collaboration}}} \bibnamefont{et~al.},
  \bibinfo{journal}{A\&A} \textbf{\bibinfo{volume}{636}}, \bibinfo{eid}{L5}
  (\bibinfo{year}{2020}).

\bibitem[{\citenamefont{{Event Horizon Telescope Collaboration}
  et~al.}(2022)}]{EHT2022}
\bibinfo{author}{\bibnamefont{{Event Horizon Telescope Collaboration}}}
  \bibnamefont{et~al.}, \bibinfo{journal}{ApJL} \textbf{\bibinfo{volume}{930}},
  \bibinfo{eid}{L12} (\bibinfo{year}{2022}).

\bibitem[{\citenamefont{{Rauch} and {Tremaine}}(1996)}]{Rauch+1996}
\bibinfo{author}{\bibfnamefont{K.~P.} \bibnamefont{{Rauch}}} \bibnamefont{and}
  \bibinfo{author}{\bibfnamefont{S.}~\bibnamefont{{Tremaine}}},
  \bibinfo{journal}{New Astron.} \textbf{\bibinfo{volume}{1}},
  \bibinfo{pages}{149} (\bibinfo{year}{1996}).

\bibitem[{\citenamefont{{Merritt}}(2013)}]{Merritt2013}
\bibinfo{author}{\bibfnamefont{D.}~\bibnamefont{{Merritt}}},
  \emph{\bibinfo{title}{{Dynamics and Evolution of Galactic Nuclei}}}
  (\bibinfo{publisher}{Princeton Univ. Press}, \bibinfo{year}{2013}).

\bibitem[{\citenamefont{{Alexander}}(2017)}]{Alexander2017}
\bibinfo{author}{\bibfnamefont{T.}~\bibnamefont{{Alexander}}},
  \bibinfo{journal}{ARA\&A} \textbf{\bibinfo{volume}{55}}, \bibinfo{pages}{17}
  (\bibinfo{year}{2017}).

\bibitem[{\citenamefont{{G{\"u}rkan} and {Hopman}}(2007)}]{Gurkan+2007}
\bibinfo{author}{\bibfnamefont{M.~A.} \bibnamefont{{G{\"u}rkan}}}
  \bibnamefont{and} \bibinfo{author}{\bibfnamefont{C.}~\bibnamefont{{Hopman}}},
  \bibinfo{journal}{MNRAS} \textbf{\bibinfo{volume}{379}},
  \bibinfo{pages}{1083} (\bibinfo{year}{2007}).

\bibitem[{\citenamefont{{Eilon} et~al.}(2009)\citenamefont{{Eilon}, {Kupi}, and
  {Alexander}}}]{Eilon+2009}
\bibinfo{author}{\bibfnamefont{E.}~\bibnamefont{{Eilon}}},
  \bibinfo{author}{\bibfnamefont{G.}~\bibnamefont{{Kupi}}}, \bibnamefont{and}
  \bibinfo{author}{\bibfnamefont{T.}~\bibnamefont{{Alexander}}},
  \bibinfo{journal}{ApJ} \textbf{\bibinfo{volume}{698}}, \bibinfo{pages}{641}
  (\bibinfo{year}{2009}).

\bibitem[{\citenamefont{{Kocsis} and {Tremaine}}(2015)}]{Kocsis+2015}
\bibinfo{author}{\bibfnamefont{B.}~\bibnamefont{{Kocsis}}} \bibnamefont{and}
  \bibinfo{author}{\bibfnamefont{S.}~\bibnamefont{{Tremaine}}},
  \bibinfo{journal}{MNRAS} \textbf{\bibinfo{volume}{448}},
  \bibinfo{pages}{3265} (\bibinfo{year}{2015}).

\bibitem[{\citenamefont{{Kocsis} and {Tremaine}}(2011)}]{Kocsis+2011}
\bibinfo{author}{\bibfnamefont{B.}~\bibnamefont{{Kocsis}}} \bibnamefont{and}
  \bibinfo{author}{\bibfnamefont{S.}~\bibnamefont{{Tremaine}}},
  \bibinfo{journal}{MNRAS} \textbf{\bibinfo{volume}{412}}, \bibinfo{pages}{187}
  (\bibinfo{year}{2011}).

\bibitem[{\citenamefont{{Hamers} et~al.}(2018)\citenamefont{{Hamers}, {Bar-Or},
  {Petrovich}, and {Antonini}}}]{Hamers+2018}
\bibinfo{author}{\bibfnamefont{A.~S.} \bibnamefont{{Hamers}}},
  \bibinfo{author}{\bibfnamefont{B.}~\bibnamefont{{Bar-Or}}},
  \bibinfo{author}{\bibfnamefont{C.}~\bibnamefont{{Petrovich}}},
  \bibnamefont{and}
  \bibinfo{author}{\bibfnamefont{F.}~\bibnamefont{{Antonini}}},
  \bibinfo{journal}{ApJ} \textbf{\bibinfo{volume}{865}}, \bibinfo{eid}{2}
  (\bibinfo{year}{2018}).

\bibitem[{\citenamefont{{Sz{\"o}lgy{\'e}n} and {Kocsis}}(2018)}]{Szolgyen+2018}
\bibinfo{author}{\bibfnamefont{{\'A}.}~\bibnamefont{{Sz{\"o}lgy{\'e}n}}}
  \bibnamefont{and} \bibinfo{author}{\bibfnamefont{B.}~\bibnamefont{{Kocsis}}},
  \bibinfo{journal}{PRL} \textbf{\bibinfo{volume}{121}}, \bibinfo{eid}{101101}
  (\bibinfo{year}{2018}).

\bibitem[{\citenamefont{{Nicholson}}(1992)}]{Nicholson1992}
\bibinfo{author}{\bibfnamefont{D.~R.} \bibnamefont{{Nicholson}}},
  \emph{\bibinfo{title}{{Introduction to Plasma Theory}}}
  (\bibinfo{publisher}{Krieger}, \bibinfo{year}{1992}).

\bibitem[{\citenamefont{{Binney} and {Tremaine}}(2008)}]{Binney+2008}
\bibinfo{author}{\bibfnamefont{J.}~\bibnamefont{{Binney}}} \bibnamefont{and}
  \bibinfo{author}{\bibfnamefont{S.}~\bibnamefont{{Tremaine}}},
  \emph{\bibinfo{title}{{Galactic Dynamics: Second Edition}}}
  (\bibinfo{publisher}{Princeton Univ. Press}, \bibinfo{year}{2008}).

\bibitem[{\citenamefont{Campa et~al.}(2014)\citenamefont{Campa, Dauxois,
  Fanelli, and Ruffo}}]{Campa+2014}
\bibinfo{author}{\bibfnamefont{A.}~\bibnamefont{Campa}},
  \bibinfo{author}{\bibfnamefont{T.}~\bibnamefont{Dauxois}},
  \bibinfo{author}{\bibfnamefont{D.}~\bibnamefont{Fanelli}}, \bibnamefont{and}
  \bibinfo{author}{\bibfnamefont{S.}~\bibnamefont{Ruffo}},
  \emph{\bibinfo{title}{Physics of Long-Range Interacting Systems}}
  (\bibinfo{publisher}{Oxford Univ. Press}, \bibinfo{year}{2014}).

\bibitem[{\citenamefont{{Gupta} and {Mukamel}}(2011)}]{Gupta+2011}
\bibinfo{author}{\bibfnamefont{S.}~\bibnamefont{{Gupta}}} \bibnamefont{and}
  \bibinfo{author}{\bibfnamefont{D.}~\bibnamefont{{Mukamel}}},
  \bibinfo{journal}{J. Stat. Mech.} \textbf{\bibinfo{volume}{2011}},
  \bibinfo{pages}{03015} (\bibinfo{year}{2011}).

\bibitem[{\citenamefont{{Maier} and {Saupe}}(1958)}]{Maier+1958}
\bibinfo{author}{\bibfnamefont{W.}~\bibnamefont{{Maier}}} \bibnamefont{and}
  \bibinfo{author}{\bibfnamefont{A.}~\bibnamefont{{Saupe}}},
  \bibinfo{journal}{Z. Nat. A.} \textbf{\bibinfo{volume}{13}},
  \bibinfo{pages}{564} (\bibinfo{year}{1958}).

\bibitem[{\citenamefont{{Roupas} et~al.}(2017)\citenamefont{{Roupas}, {Kocsis},
  and {Tremaine}}}]{Roupas+2017}
\bibinfo{author}{\bibfnamefont{Z.}~\bibnamefont{{Roupas}}},
  \bibinfo{author}{\bibfnamefont{B.}~\bibnamefont{{Kocsis}}}, \bibnamefont{and}
  \bibinfo{author}{\bibfnamefont{S.}~\bibnamefont{{Tremaine}}},
  \bibinfo{journal}{ApJ} \textbf{\bibinfo{volume}{842}}, \bibinfo{eid}{90}
  (\bibinfo{year}{2017}).

\bibitem[{\citenamefont{{Roupas}}(2020)}]{Roupas2020}
\bibinfo{author}{\bibfnamefont{Z.}~\bibnamefont{{Roupas}}},
  \bibinfo{journal}{J. Phys. A} \textbf{\bibinfo{volume}{53}},
  \bibinfo{eid}{045002} (\bibinfo{year}{2020}).

\bibitem[{\citenamefont{{Taylor} and {McNamara}}(1971)}]{Taylor+1971}
\bibinfo{author}{\bibfnamefont{J.~B.} \bibnamefont{{Taylor}}} \bibnamefont{and}
  \bibinfo{author}{\bibfnamefont{B.}~\bibnamefont{{McNamara}}},
  \bibinfo{journal}{Phys. Fluids} \textbf{\bibinfo{volume}{14}},
  \bibinfo{pages}{1492} (\bibinfo{year}{1971}).

\bibitem[{\citenamefont{{Chavanis} et~al.}(1996)\citenamefont{{Chavanis},
  {Sommeria}, and {Robert}}}]{Chavanis+1996}
\bibinfo{author}{\bibfnamefont{P.-H.} \bibnamefont{{Chavanis}}},
  \bibinfo{author}{\bibfnamefont{J.}~\bibnamefont{{Sommeria}}},
  \bibnamefont{and} \bibinfo{author}{\bibfnamefont{R.}~\bibnamefont{{Robert}}},
  \bibinfo{journal}{ApJ} \textbf{\bibinfo{volume}{471}}, \bibinfo{pages}{385}
  (\bibinfo{year}{1996}).

\bibitem[{\citenamefont{{Garc\`{i}a-Ojalvo} and
  {Sancho}}(1999)}]{GarciaOjalvo+1999}
\bibinfo{author}{\bibfnamefont{J.}~\bibnamefont{{Garc\`{i}a-Ojalvo}}}
  \bibnamefont{and} \bibinfo{author}{\bibfnamefont{J.~M.}
  \bibnamefont{{Sancho}}}, \emph{\bibinfo{title}{{Noise in Spatially Extended
  Systems }}} (\bibinfo{publisher}{Springer}, \bibinfo{year}{1999}).

\bibitem[{\citenamefont{{Krommes}}(2002)}]{Krommes2002}
\bibinfo{author}{\bibfnamefont{J.~A.} \bibnamefont{{Krommes}}},
  \bibinfo{journal}{Phys. Rep.} \textbf{\bibinfo{volume}{360}},
  \bibinfo{pages}{1} (\bibinfo{year}{2002}).

\bibitem[{\citenamefont{{Zhou}}(2021)}]{Zhou2021}
\bibinfo{author}{\bibfnamefont{Y.}~\bibnamefont{{Zhou}}},
  \bibinfo{journal}{Phys. Rep.} \textbf{\bibinfo{volume}{935}},
  \bibinfo{pages}{1} (\bibinfo{year}{2021}).

\bibitem[{\citenamefont{{Kraichnan}}(1959{\natexlab{a}})}]{Kraichnan1959}
\bibinfo{author}{\bibfnamefont{R.~H.} \bibnamefont{{Kraichnan}}},
  \bibinfo{journal}{J. Fluid Mech.} \textbf{\bibinfo{volume}{5}},
  \bibinfo{pages}{497} (\bibinfo{year}{1959}{\natexlab{a}}).

\bibitem[{\citenamefont{{Martin} et~al.}(1973)\citenamefont{{Martin}, {Siggia},
  and {Rose}}}]{MSR1973}
\bibinfo{author}{\bibfnamefont{P.~C.} \bibnamefont{{Martin}}},
  \bibinfo{author}{\bibfnamefont{E.~D.} \bibnamefont{{Siggia}}},
  \bibnamefont{and} \bibinfo{author}{\bibfnamefont{H.~A.}
  \bibnamefont{{Rose}}}, \bibinfo{journal}{Phys. Rev. A}
  \textbf{\bibinfo{volume}{8}}, \bibinfo{pages}{423} (\bibinfo{year}{1973}).

\bibitem[{\citenamefont{{Magnan} et~al.}(2022)}]{Magnan+2022}
\bibinfo{author}{\bibfnamefont{N.}~\bibnamefont{{Magnan}}}
  \bibnamefont{et~al.}, \bibinfo{journal}{MNRAS}
  \textbf{\bibinfo{volume}{514}}, \bibinfo{pages}{3452} (\bibinfo{year}{2022}).

\bibitem[{\citenamefont{{Fouvry} et~al.}(2019)\citenamefont{{Fouvry}, {Bar-Or},
  and {Chavanis}}}]{Fouvry+2019}
\bibinfo{author}{\bibfnamefont{J.-B.} \bibnamefont{{Fouvry}}},
  \bibinfo{author}{\bibfnamefont{B.}~\bibnamefont{{Bar-Or}}}, \bibnamefont{and}
  \bibinfo{author}{\bibfnamefont{P.-H.} \bibnamefont{{Chavanis}}},
  \bibinfo{journal}{ApJ} \textbf{\bibinfo{volume}{883}}, \bibinfo{eid}{161}
  (\bibinfo{year}{2019}).

\bibitem[{\citenamefont{{Berera} et~al.}(2013)\citenamefont{{Berera},
  {Salewski}, and {McComb}}}]{Berera+2013}
\bibinfo{author}{\bibfnamefont{A.}~\bibnamefont{{Berera}}},
  \bibinfo{author}{\bibfnamefont{M.}~\bibnamefont{{Salewski}}},
  \bibnamefont{and} \bibinfo{author}{\bibfnamefont{W.~D.}
  \bibnamefont{{McComb}}}, \bibinfo{journal}{Phys. Rev. E}
  \textbf{\bibinfo{volume}{87}}, \bibinfo{eid}{013007} (\bibinfo{year}{2013}).

\bibitem[{\citenamefont{{Krommes}}(2015)}]{Krommes2015}
\bibinfo{author}{\bibfnamefont{J.~A.} \bibnamefont{{Krommes}}},
  \bibinfo{journal}{JPP} \textbf{\bibinfo{volume}{81}},
  \bibinfo{eid}{205810601} (\bibinfo{year}{2015}).

\bibitem[{\citenamefont{Bernardeau et~al.}(2012)\citenamefont{Bernardeau,
  Van~de Rijt, and Vernizzi}}]{Bernardeau+2012}
\bibinfo{author}{\bibfnamefont{F.}~\bibnamefont{Bernardeau}},
  \bibinfo{author}{\bibfnamefont{N.}~\bibnamefont{Van~de Rijt}},
  \bibnamefont{and} \bibinfo{author}{\bibfnamefont{F.}~\bibnamefont{Vernizzi}},
  \bibinfo{journal}{Phys. Rev. D} \textbf{\bibinfo{volume}{85}},
  \bibinfo{pages}{063509} (\bibinfo{year}{2012}).

\bibitem[{\citenamefont{Lancaster and Blundell}(2014)}]{Lancaster+2014}
\bibinfo{author}{\bibfnamefont{T.}~\bibnamefont{Lancaster}} \bibnamefont{and}
  \bibinfo{author}{\bibfnamefont{S.}~\bibnamefont{Blundell}},
  \emph{\bibinfo{title}{Quantum Field Theory for the Gifted Amateur}}
  (\bibinfo{publisher}{OUP Oxford}, \bibinfo{year}{2014}).

\bibitem[{\citenamefont{{Peskin} and {Schroeder}}(2018)}]{Peskin+2018}
\bibinfo{author}{\bibfnamefont{M.~E.} \bibnamefont{{Peskin}}} \bibnamefont{and}
  \bibinfo{author}{\bibfnamefont{D.~V.} \bibnamefont{{Schroeder}}},
  \emph{\bibinfo{title}{An Introduction To Quantum Field Theory}}
  (\bibinfo{publisher}{CRC Press}, \bibinfo{year}{2018}).

\bibitem[{\citenamefont{{Rose}}(1985)}]{Rose1985}
\bibinfo{author}{\bibfnamefont{H.~A.} \bibnamefont{{Rose}}},
  \bibinfo{journal}{Physica D} \textbf{\bibinfo{volume}{14}},
  \bibinfo{pages}{216} (\bibinfo{year}{1985}).

\bibitem[{\citenamefont{{Krommes}}(1984)}]{Krommes1984}
\bibinfo{author}{\bibfnamefont{J.~A.} \bibnamefont{{Krommes}}}, in
  \emph{\bibinfo{booktitle}{Basic Plasma Physics, Volume 1}}
  (\bibinfo{year}{1984}), p. \bibinfo{pages}{183}.

\bibitem[{\citenamefont{{Kraichnan}}(1958)}]{Kraichnan1958}
\bibinfo{author}{\bibfnamefont{R.~H.} \bibnamefont{{Kraichnan}}},
  \bibinfo{journal}{Phys. Rev.} \textbf{\bibinfo{volume}{109}},
  \bibinfo{pages}{1407} (\bibinfo{year}{1958}).

\bibitem[{\citenamefont{{Ottaviani}}(1990)}]{Ottaviani1990}
\bibinfo{author}{\bibfnamefont{M.}~\bibnamefont{{Ottaviani}}},
  \bibinfo{journal}{Phys. Lett. A} \textbf{\bibinfo{volume}{143}},
  \bibinfo{pages}{325} (\bibinfo{year}{1990}).

\bibitem[{\citenamefont{{Kolmogorov}}(1941)}]{Kolmogorov1941}
\bibinfo{author}{\bibfnamefont{A.}~\bibnamefont{{Kolmogorov}}},
  \bibinfo{journal}{Acad. Sci. USSR} \textbf{\bibinfo{volume}{30}},
  \bibinfo{pages}{301} (\bibinfo{year}{1941}).

\bibitem[{\citenamefont{{Kraichnan}}(1964)}]{Kraichnan1964}
\bibinfo{author}{\bibfnamefont{R.~H.} \bibnamefont{{Kraichnan}}},
  \bibinfo{journal}{Physics of Fluids} \textbf{\bibinfo{volume}{7}},
  \bibinfo{pages}{1717} (\bibinfo{year}{1964}).

\bibitem[{\citenamefont{{Kraichnan}}(1965)}]{Kraichnan1965}
\bibinfo{author}{\bibfnamefont{R.~H.} \bibnamefont{{Kraichnan}}},
  \bibinfo{journal}{Physics of Fluids} \textbf{\bibinfo{volume}{8}},
  \bibinfo{pages}{575} (\bibinfo{year}{1965}).

\bibitem[{\citenamefont{{McComb}}(1990)}]{McComb1990}
\bibinfo{author}{\bibfnamefont{W.~D.} \bibnamefont{{McComb}}},
  \emph{\bibinfo{title}{{The physics of fluid turbulence}}}
  (\bibinfo{publisher}{Clarendon Press}, \bibinfo{year}{1990}).

\bibitem[{\citenamefont{Frisch}(1995)}]{Frisch1995}
\bibinfo{author}{\bibfnamefont{U.}~\bibnamefont{Frisch}},
  \emph{\bibinfo{title}{Turbulence: The Legacy of A. N. Kolmogorov}}
  (\bibinfo{publisher}{Cambridge University Press}, \bibinfo{year}{1995}).

\bibitem[{\citenamefont{{Kraichnan}}(1959{\natexlab{b}})}]{Kraichnan1959a}
\bibinfo{author}{\bibfnamefont{R.~H.} \bibnamefont{{Kraichnan}}},
  \bibinfo{journal}{Phys. Rev.} \textbf{\bibinfo{volume}{113}},
  \bibinfo{pages}{1181} (\bibinfo{year}{1959}{\natexlab{b}}).

\bibitem[{\citenamefont{{Kraichnan}}(1961)}]{Kraichnan1961}
\bibinfo{author}{\bibfnamefont{R.~H.} \bibnamefont{{Kraichnan}}},
  \bibinfo{journal}{J. Math. Phys.} \textbf{\bibinfo{volume}{2}},
  \bibinfo{pages}{124} (\bibinfo{year}{1961}).

\bibitem[{\citenamefont{{Valageas}}(2004)}]{Valageas2004}
\bibinfo{author}{\bibfnamefont{P.}~\bibnamefont{{Valageas}}},
  \bibinfo{journal}{A\&A} \textbf{\bibinfo{volume}{421}}, \bibinfo{pages}{23}
  (\bibinfo{year}{2004}).

\bibitem[{\citenamefont{{Fukuda} et~al.}(1995)}]{Fukuda1995}
\bibinfo{author}{\bibfnamefont{R.}~\bibnamefont{{Fukuda}}}
  \bibnamefont{et~al.}, \bibinfo{journal}{Prog. Theor. Phys.}
  \textbf{\bibinfo{volume}{121}}, \bibinfo{pages}{1} (\bibinfo{year}{1995}).

\bibitem[{\citenamefont{{Cornwall} et~al.}(1974)\citenamefont{{Cornwall},
  {Jackiw}, and {Tomboulis}}}]{Cornwall1974}
\bibinfo{author}{\bibfnamefont{J.~M.} \bibnamefont{{Cornwall}}},
  \bibinfo{author}{\bibfnamefont{R.}~\bibnamefont{{Jackiw}}}, \bibnamefont{and}
  \bibinfo{author}{\bibfnamefont{E.}~\bibnamefont{{Tomboulis}}},
  \bibinfo{journal}{Phys. Rev. D} \textbf{\bibinfo{volume}{10}},
  \bibinfo{pages}{2428} (\bibinfo{year}{1974}).

\bibitem[{\citenamefont{{Berges}}(2004)}]{Berges2004}
\bibinfo{author}{\bibfnamefont{J.}~\bibnamefont{{Berges}}},
  \bibinfo{journal}{\prd} \textbf{\bibinfo{volume}{70}}, \bibinfo{eid}{105010}
  (\bibinfo{year}{2004}).

\bibitem[{\citenamefont{{Carrington} and {Guo}}(2011)}]{Carrington2011}
\bibinfo{author}{\bibfnamefont{M.~E.} \bibnamefont{{Carrington}}}
  \bibnamefont{and} \bibinfo{author}{\bibfnamefont{Y.}~\bibnamefont{{Guo}}},
  \bibinfo{journal}{Phys. Rev. D} \textbf{\bibinfo{volume}{83}},
  \bibinfo{eid}{016006} (\bibinfo{year}{2011}).

\bibitem[{\citenamefont{{Zhou}}(2010)}]{Zhou2010}
\bibinfo{author}{\bibfnamefont{Y.}~\bibnamefont{{Zhou}}},
  \bibinfo{journal}{Phys. Rep.} \textbf{\bibinfo{volume}{488}},
  \bibinfo{pages}{1} (\bibinfo{year}{2010}).

\bibitem[{\citenamefont{{Delamotte}}(2012)}]{Delamotte2012}
\bibinfo{author}{\bibfnamefont{B.}~\bibnamefont{{Delamotte}}}, in
  \emph{\bibinfo{booktitle}{Lecture Notes in Physics}}
  (\bibinfo{publisher}{Springer}, \bibinfo{year}{2012}), vol.
  \bibinfo{volume}{852}, p.~\bibinfo{pages}{49}.

\bibitem[{\citenamefont{{Dupuis} et~al.}(2021)}]{Dupuis+2021}
\bibinfo{author}{\bibfnamefont{N.}~\bibnamefont{{Dupuis}}}
  \bibnamefont{et~al.}, \bibinfo{journal}{Phys. Rep.}
  \textbf{\bibinfo{volume}{910}}, \bibinfo{pages}{1} (\bibinfo{year}{2021}).

\bibitem[{\citenamefont{{Canet}}(2022)}]{Canet2022}
\bibinfo{author}{\bibfnamefont{L.}~\bibnamefont{{Canet}}}, \bibinfo{journal}{J.
  Fluid. Mech} \textbf{\bibinfo{volume}{950}}, \bibinfo{eid}{P1}
  (\bibinfo{year}{2022}), \eprint{2205.01427}.

\bibitem[{\citenamefont{{Fontaine} et~al.}(2023)\citenamefont{{Fontaine},
  {Tarpin}, {Bouchet}, and {Canet}}}]{Fontaine+2023}
\bibinfo{author}{\bibfnamefont{C.}~\bibnamefont{{Fontaine}}},
  \bibinfo{author}{\bibfnamefont{M.}~\bibnamefont{{Tarpin}}},
  \bibinfo{author}{\bibfnamefont{F.}~\bibnamefont{{Bouchet}}},
  \bibnamefont{and} \bibinfo{author}{\bibfnamefont{L.}~\bibnamefont{{Canet}}},
  \bibinfo{journal}{SciPost Physics} \textbf{\bibinfo{volume}{15}},
  \bibinfo{eid}{212} (\bibinfo{year}{2023}), \eprint{2208.00225}.

\bibitem[{\citenamefont{{Tarpin} et~al.}(2019)}]{Tarpin+2019}
\bibinfo{author}{\bibfnamefont{M.}~\bibnamefont{{Tarpin}}}
  \bibnamefont{et~al.}, \bibinfo{journal}{J. Phys. A}
  \textbf{\bibinfo{volume}{52}}, \bibinfo{eid}{085501} (\bibinfo{year}{2019}).

\bibitem[{\citenamefont{{Touma} et~al.}(2019)\citenamefont{{Touma}, {Tremaine},
  and {Kazandjian}}}]{Touma+2019}
\bibinfo{author}{\bibfnamefont{J.}~\bibnamefont{{Touma}}},
  \bibinfo{author}{\bibfnamefont{S.}~\bibnamefont{{Tremaine}}},
  \bibnamefont{and}
  \bibinfo{author}{\bibfnamefont{M.}~\bibnamefont{{Kazandjian}}},
  \bibinfo{journal}{PRL} \textbf{\bibinfo{volume}{123}}, \bibinfo{eid}{021103}
  (\bibinfo{year}{2019}).

\bibitem[{\citenamefont{{Gruzinov} et~al.}(2020)\citenamefont{{Gruzinov},
  {Levin}, and {Zhu}}}]{Gruzinov+2020}
\bibinfo{author}{\bibfnamefont{A.}~\bibnamefont{{Gruzinov}}},
  \bibinfo{author}{\bibfnamefont{Y.}~\bibnamefont{{Levin}}}, \bibnamefont{and}
  \bibinfo{author}{\bibfnamefont{J.}~\bibnamefont{{Zhu}}},
  \bibinfo{journal}{ApJ} \textbf{\bibinfo{volume}{905}}, \bibinfo{eid}{11}
  (\bibinfo{year}{2020}).

\bibitem[{\citenamefont{{M{\'a}th{\'e}}
  et~al.}(2023)\citenamefont{{M{\'a}th{\'e}}, {Sz{\"o}lgy{\'e}n}, and
  {Kocsis}}}]{Mathe+2023}
\bibinfo{author}{\bibfnamefont{G.}~\bibnamefont{{M{\'a}th{\'e}}}},
  \bibinfo{author}{\bibfnamefont{{\'A}.}~\bibnamefont{{Sz{\"o}lgy{\'e}n}}},
  \bibnamefont{and} \bibinfo{author}{\bibfnamefont{B.}~\bibnamefont{{Kocsis}}},
  \bibinfo{journal}{MNRAS} \textbf{\bibinfo{volume}{520}},
  \bibinfo{pages}{2204} (\bibinfo{year}{2023}).

\bibitem[{\citenamefont{{Tak{\'a}cs} and {Kocsis}}(2018)}]{Takacs+2018}
\bibinfo{author}{\bibfnamefont{{\'A}.}~\bibnamefont{{Tak{\'a}cs}}}
  \bibnamefont{and} \bibinfo{author}{\bibfnamefont{B.}~\bibnamefont{{Kocsis}}},
  \bibinfo{journal}{ApJ} \textbf{\bibinfo{volume}{856}}, \bibinfo{eid}{113}
  (\bibinfo{year}{2018}).

\bibitem[{\citenamefont{{Fragione} and {Loeb}}(2022)}]{Fragione+2022}
\bibinfo{author}{\bibfnamefont{G.}~\bibnamefont{{Fragione}}} \bibnamefont{and}
  \bibinfo{author}{\bibfnamefont{A.}~\bibnamefont{{Loeb}}},
  \bibinfo{journal}{ApJL} \textbf{\bibinfo{volume}{932}}, \bibinfo{eid}{L17}
  (\bibinfo{year}{2022}).

\bibitem[{\citenamefont{{GRAVITY Collaboration} et~al.}(2023)}]{Gravity+2023}
\bibinfo{author}{\bibnamefont{{GRAVITY Collaboration}}} \bibnamefont{et~al.},
  \bibinfo{journal}{A\&A} \textbf{\bibinfo{volume}{672}}, \bibinfo{eid}{A63}
  (\bibinfo{year}{2023}).

\bibitem[{\citenamefont{{Will} et~al.}(2023)}]{Will+2023}
\bibinfo{author}{\bibfnamefont{C.~M.} \bibnamefont{{Will}}}
  \bibnamefont{et~al.}, \bibinfo{journal}{ApJ} \textbf{\bibinfo{volume}{959}},
  \bibinfo{eid}{58} (\bibinfo{year}{2023}).

\bibitem[{\citenamefont{{Ginat} et~al.}(2023)\citenamefont{{Ginat},
  {Panamarev}, {Kocsis}, and {Perets}}}]{Ginat+2022}
\bibinfo{author}{\bibfnamefont{Y.~B.} \bibnamefont{{Ginat}}},
  \bibinfo{author}{\bibfnamefont{T.}~\bibnamefont{{Panamarev}}},
  \bibinfo{author}{\bibfnamefont{B.}~\bibnamefont{{Kocsis}}}, \bibnamefont{and}
  \bibinfo{author}{\bibfnamefont{H.~B.} \bibnamefont{{Perets}}},
  \bibinfo{journal}{MNRAS} \textbf{\bibinfo{volume}{525}},
  \bibinfo{pages}{4202} (\bibinfo{year}{2023}).

\bibitem[{\citenamefont{{Giral Mart{\'\i}nez} et~al.}(2020)\citenamefont{{Giral
  Mart{\'\i}nez}, {Fouvry}, and {Pichon}}}]{Giral+2020}
\bibinfo{author}{\bibfnamefont{J.}~\bibnamefont{{Giral Mart{\'\i}nez}}},
  \bibinfo{author}{\bibfnamefont{J.-B.} \bibnamefont{{Fouvry}}},
  \bibnamefont{and} \bibinfo{author}{\bibfnamefont{C.}~\bibnamefont{{Pichon}}},
  \bibinfo{journal}{MNRAS} \textbf{\bibinfo{volume}{499}},
  \bibinfo{pages}{2714} (\bibinfo{year}{2020}).

\bibitem[{\citenamefont{{Bartko} et~al.}(2009)}]{Bartko+2009}
\bibinfo{author}{\bibfnamefont{H.}~\bibnamefont{{Bartko}}}
  \bibnamefont{et~al.}, \bibinfo{journal}{ApJ} \textbf{\bibinfo{volume}{697}},
  \bibinfo{pages}{1741} (\bibinfo{year}{2009}).

\bibitem[{\citenamefont{{Lu} et~al.}(2009)}]{Lu+2009}
\bibinfo{author}{\bibfnamefont{J.~R.} \bibnamefont{{Lu}}} \bibnamefont{et~al.},
  \bibinfo{journal}{ApJ} \textbf{\bibinfo{volume}{690}}, \bibinfo{pages}{1463}
  (\bibinfo{year}{2009}).

\bibitem[{\citenamefont{{Yelda} et~al.}(2014)}]{Yelda+2014}
\bibinfo{author}{\bibfnamefont{S.}~\bibnamefont{{Yelda}}} \bibnamefont{et~al.},
  \bibinfo{journal}{ApJ} \textbf{\bibinfo{volume}{783}}, \bibinfo{eid}{131}
  (\bibinfo{year}{2014}).

\bibitem[{\citenamefont{{von Fellenberg} et~al.}(2022)}]{vonFellenberg+2022}
\bibinfo{author}{\bibfnamefont{S.~D.} \bibnamefont{{von Fellenberg}}}
  \bibnamefont{et~al.}, \bibinfo{journal}{ApJL} \textbf{\bibinfo{volume}{932}},
  \bibinfo{eid}{L6} (\bibinfo{year}{2022}).

\bibitem[{\citenamefont{{Panamarev} and {Kocsis}}(2022)}]{Panamarev+2022}
\bibinfo{author}{\bibfnamefont{T.}~\bibnamefont{{Panamarev}}} \bibnamefont{and}
  \bibinfo{author}{\bibfnamefont{B.}~\bibnamefont{{Kocsis}}},
  \bibinfo{journal}{MNRAS} \textbf{\bibinfo{volume}{517}},
  \bibinfo{pages}{6205} (\bibinfo{year}{2022}).

\bibitem[{\citenamefont{{Fouvry} et~al.}(2023)\citenamefont{{Fouvry},
  {Bustamante-Rosell}, and {Zimmerman}}}]{Fouvry+2023}
\bibinfo{author}{\bibfnamefont{J.-B.} \bibnamefont{{Fouvry}}},
  \bibinfo{author}{\bibfnamefont{M.~J.} \bibnamefont{{Bustamante-Rosell}}},
  \bibnamefont{and}
  \bibinfo{author}{\bibfnamefont{A.}~\bibnamefont{{Zimmerman}}},
  \bibinfo{journal}{MNRAS} \textbf{\bibinfo{volume}{526}},
  \bibinfo{pages}{1471} (\bibinfo{year}{2023}).

\bibitem[{\citenamefont{{Varshalovich} et~al.}(1988)}]{Varshalovich1988}
\bibinfo{author}{\bibfnamefont{D.~A.} \bibnamefont{{Varshalovich}}}
  \bibnamefont{et~al.}, \emph{\bibinfo{title}{Quantum Theory of Angular
  Momentum}} (\bibinfo{publisher}{World Scientific}, \bibinfo{year}{1988}).

\bibitem[{\citenamefont{Binney and Skinner}(2013)}]{Binney+2013}
\bibinfo{author}{\bibfnamefont{J.}~\bibnamefont{Binney}} \bibnamefont{and}
  \bibinfo{author}{\bibfnamefont{D.}~\bibnamefont{Skinner}},
  \emph{\bibinfo{title}{The Physics of Quantum Mechanics}}
  (\bibinfo{publisher}{OUP Oxford}, \bibinfo{year}{2013}).

\bibitem[{\citenamefont{{Fouvry} et~al.}(2022)\citenamefont{{Fouvry}, {Dehnen},
  {Tremaine}, and {Bar-Or}}}]{Fouvry+2022}
\bibinfo{author}{\bibfnamefont{J.-B.} \bibnamefont{{Fouvry}}},
  \bibinfo{author}{\bibfnamefont{W.}~\bibnamefont{{Dehnen}}},
  \bibinfo{author}{\bibfnamefont{S.}~\bibnamefont{{Tremaine}}},
  \bibnamefont{and} \bibinfo{author}{\bibfnamefont{B.}~\bibnamefont{{Bar-Or}}},
  \bibinfo{journal}{ApJ} \textbf{\bibinfo{volume}{931}}, \bibinfo{eid}{8}
  (\bibinfo{year}{2022}).

\bibitem[{git()}]{github}
\bibinfo{howpublished}{\url{https://github.com/sfloresmo/VRR_DIA}}.

\end{thebibliography}
\end{document}